\documentclass[a4paper,english,aps,reprint]{revtex4-1}
\usepackage[T1]{fontenc}
\usepackage[latin9]{inputenc}
\usepackage{babel}
\usepackage{verbatim}
\usepackage{float}
\usepackage{amsmath}
\usepackage{amssymb}
\usepackage{stackrel}
\usepackage{graphicx}
\usepackage{esint}
\usepackage[unicode=true,pdfusetitle,
 bookmarks=true,bookmarksnumbered=false,bookmarksopen=true,bookmarksopenlevel=1,
 breaklinks=false,pdfborder={0 0 0},backref=false,colorlinks=false]
 {hyperref}

\makeatletter

\pdfpageheight\paperheight
\pdfpagewidth\paperwidth

\newcommand{\lyxdot}{.}

\@ifundefined{textcolor}{}
{%
 \definecolor{BLACK}{gray}{0}
 \definecolor{WHITE}{gray}{1}
 \definecolor{RED}{rgb}{1,0,0}
 \definecolor{GREEN}{rgb}{0,1,0}
 \definecolor{BLUE}{rgb}{0,0,1}
 \definecolor{CYAN}{cmyk}{1,0,0,0}
 \definecolor{MAGENTA}{cmyk}{0,1,0,0}
 \definecolor{YELLOW}{cmyk}{0,0,1,0}
}

\usepackage{babel}
\usepackage{babel}
\usepackage[normalem]{ulem}
\usepackage{babel}
\usepackage{babel}
\usepackage{amsfonts}
\providecommand{\U}[1]{\protect\rule{.1in}{.1in}}
\pdfpageheight\paperheight
\pdfpagewidth\paperwidth

\@ifundefined{textcolor}{}{
\definecolor{BLACK}{gray}{0}
\definecolor{WHITE}{gray}{1}
\definecolor{RED}{rgb}{1,0,0}
\definecolor{GREEN}{rgb}{0,1,0}
\definecolor{BLUE}{rgb}{0,0,1}
\definecolor{CYAN}{cmyk}{1,0,0,0}
\definecolor{MAGENTA}{cmyk}{0,1,0,0}
\definecolor{YELLOW}{cmyk}{0,0,1,0}
}

\makeatother

\begin{document}

\title{Quantum entanglement in strong-field ionization}

\author{Szilárd Majorosi}

\email{majorosi.szilard@physx.u-szeged.hu}

\affiliation{Department of Theoretical Physics, University of Szeged\linebreak{}
Tisza L. krt. 84-86, H-6720 Szeged}

\author{Mihály G. Benedict}

\email{benedict@physx.u-szeged.hu}

\affiliation{Department of Theoretical Physics, University of Szeged\linebreak{}
Tisza L. krt. 84-86, H-6720 Szeged}

\author{Attila Czirják}

\email{czirjak@physx.u-szeged.hu}

\affiliation{Department of Theoretical Physics, University of Szeged\linebreak{}
Tisza L. krt. 84-86, H-6720 Szeged}

\affiliation{ELI-ALPS, ELI-HU Non-Profit Ltd., Dugonics tér 13, H-6720 Szeged,
Hungary}
\begin{abstract}
We investigate the time-evolution of quantum entanglement between
an electron, liberated by a strong few-cycle laser pulse, and its
parent ion-core. Since the standard procedure is numerically prohibitive
in this case, we propose a novel way to quantify the quantum correlation
in such a system: we use the reduced density matrices of the directional
subspaces along the polarization of the laser pulse and along the
transverse directions as building blocks for an approximate entanglement
entropy. We present our results, based on accurate numerical simulations,
in terms of several of these entropies, for selected values of the
peak electric field strength and the carrier-envelope phase difference
of the laser pulse. The time evolution of the mutual entropy of the
electron and the ion-core motion along the direction of the laser
polarization is similar to our earlier results based on a simple one-dimensional
model. However, taking into account also the dynamics perpendicular
to the laser polarization reveals a surprisingly different entanglement
dynamics above the laser intensity range corresponding to pure tunneling:
the quantum entanglement decreases with time in the over-the-barrier
ionization regime. 
\end{abstract}
\maketitle

\section{Introduction\label{sec:introduction}}

Although quantum entanglement between two particles' spatial motion
(i.e. their positions or momenta) dates back to the early days of
quantum mechanics \cite{einstein1935quantumEPR,reid2009EPRparadoxall},
the features of continuous variable quantum entanglement \cite{braunstein2005entcontinuousvariables}
are still much less explored and utilized than those of discrete variables
systems \cite{horodecki2009entanglementreview}. The few-particle
quantum systems studied in connection with quantum entanglement usually
need special preparation procedures and they are typically very sensitive
to environmental circumstances. In contrast to this, the strong-field
ionization of an atom is a very well explored and understood process,
both theoretically and experimentally \cite{Keldysh_JETP_1965,kulander1987tdseionization,Ferray_JPhysB_1988_HHG,Corkum_PRL_1993,varro1993multiphotonequation,becker1994tdsehhgmodel,lewenstein1994hhgtheory,protopapas1997tdseionization,hentschel2001attosecond,kienberger2002attosecond,drescher2002attosecond,baltuska2003attosecond,ivanov2005strongfield,Uiberacker_Nature_2007,eckle2008attosecond,Krausz_RevModPhys_2009_Attosecond_physics,Schultze_Science_2010,lein2011streakingionization,balogh2011attohhg,pfeiffer2012ionizationmomentum,Shafir_Nature_2012,gombkoto2016quantoptichhg}.
However, despite the fact that it is widely used in standard procedures
to generate e.g. high-order harmonic radiation \cite{McPherson_JOSAB_1987_HHG,Harris_OptCom_1993_HHG_Atto},
it is very little known that this strong-field ionization generates
also quantum entanglement between the liberated electron and its parent
ion-core. In our earlier work \cite{czirjak2012entanglementbuild,czirjak2013rescatterentanglement},
based on a simple one-dimensional model, we have already shown that
the time evolution of this quantum entanglement shows interesting
features. A straightforward question is, whether these are also present
in the strong-field ionization of a real atom? In the present paper,
we report about our new results on this process: although we keep
the single active electron approximation, we do the investigation
in 3 spatial dimensions, using the true Coulomb potential.

Most of the papers on quantum entanglement in light induced atomic
processes study the correlations between the emitted photon and the
emitting atomic system \cite{fedorov2005entatomphoton,volz2006entatomphoton,xie2009entatomfield}.
Papers on entanglement between a charged particle and a photon \cite{varro2008entphotonelectron,varro2010entphotonelectron},
entanglement in two particles\textquoteright{} collision \cite{tal2005entcollision,benedict2012entanglementtwo,nagy2012exactentangled,kull2012electronionscatter,adivi2014entanglementgeneration,feder2015entanglementcreation}
and on the temporal change of the correlation potential during electron
tunneling from a molecule \cite{walters2010attotunneling} give valuable
insight into the quantum features of problems related to our present
paper. Entanglement between the fragments of an atomic system due
to a light-induced break-up process, like photoionization and photodissociation,
was studied by Fedorov and coworkers \citep{fedorov2004entpacketnarrowing,fedorov2006strongfieldentangled}
in the framework of Gaussian states. However, this latter approach
does not seem suitable enough to deal with the problem of quantum
entanglement during the strong-field ionization of an atom, which
motivated us to perform an accurate numerical investigation of the
problem.

This paper is organized as follows: in Section \ref{sec:strong-field},
we outline the solution of the quantum mechanical two-body Coulomb
problem under the influence of an external laser field. In Section
\ref{sec:entanglement-calculation}, we present the details of our
entanglement calculations which is based on the directionally reduced
dynamics. Amongst others, we introduce the spatial entropy of the
wave function and the correlation entropies within the directional
subspaces. Using the corresponding directional entropies, we propose
an approximate formula to quantify the total electron-core entanglement
we actually seek for. Based on this latter, we also discuss the connection
to the results based on one dimensional models. We present our numerical
results on the temporal evolution of quantum entanglement during the
strong-field ionization process in Section \ref{sec:results}. We
show how do the specific quantities, including several different entropies,
reflect the system's behavior, and we investigate in detail their
dependence on the intensity and the carrier-envelope phase difference
(CEP) of the laser pulse. Finally, we discuss the relevance of our
main results in Section \ref{sec:summary}. In the Appendix, we summarize
the necessary theoretical background for quantum entanglement between
two particles, and recall certain notions (e.g. correlation types,
quantum conditional entropy, quantum mutual entropy) which are important
for directionally reduced subsystems.

We use atomic units throughout this article (i.e. $\hbar=1$, $e=1$,
$m_{e}=1$) unless stated otherwise.

\section{Strong Field Ionization\label{sec:strong-field}}

\subsection{Two-body Hamiltonian\label{sub:strong-field-hamiltonian}}

The quantum mechanical description of a hydrogen atom, or any other
atom in the single active electron approximation, driven by a strong
laser pulse, is naturally carried out as a two-body (or bipartite)
problem consisting of the electron ($e$) - ion-core ($c$) system.
We consider their interaction with the laser pulse in the dipole approximation,
i.e. as an external time-dependent electric field, because the relevant
EM field wavelengths exceed the size of the system by several orders
of magnitude. Using the length gauge \citep{BOOK_ATOMS_IN_INTENSE_FIELDS_2012}
we have the following Hamiltonian for this system: 
\begin{equation}
H_{ec}=\frac{\mathbf{P}_{e}^{2}}{2m_{e}}+\frac{\mathbf{P}_{c}^{2}}{2m_{c}}-\frac{1}{|\mathbf{r}_{e}-\mathbf{r}_{c}|}+\mathbf{E}(t)(\mathbf{r}_{e}-\mathbf{r}_{c}),\label{eq:quant_hamiltonian_two_ec}
\end{equation}
where $m_{e}(=1)$ and $m_{c}$ are the electron and core masses,
respectively.

As it is well known, this problem can be simplified by performing
a coordinate transformation to the center of mass ($\mathbf{r}_{0},\mathbf{P}_{0}$)
and relative coordinates ($\mathbf{r},\mathbf{P}$) as 
\begin{equation}
\begin{array}{ccccccc}
\,\mathbf{r}_{0} & = & \alpha_{e}\mathbf{r}_{e}+\alpha_{c}\mathbf{r}_{c}, & \, & \,\mathbf{P}_{0} & = & \mathbf{P}_{e}+\mathbf{P}_{c},\,{\,}{\,}\\
\mathbf{r} & = & \mathbf{r}_{e}-\mathbf{r}_{c}, & \, & \mathbf{P} & = & \alpha_{c}\mathbf{P}_{e}-\alpha_{e}\mathbf{P}_{c},
\end{array}\label{eq:quant_twobody_transform}
\end{equation}
where 
\begin{equation}
\alpha_{e}=m_{e}/M,\,\,\alpha_{c}=m_{c}/M,\,\, M=m_{e}+m_{c}.\label{eq:ent_calc_dm_z_alpha}
\end{equation}
Using also the reduced mass $\mu=m_{e}m_{c}/M$, we obtain the Hamiltonian
\begin{equation}
H_{ec}=\frac{\mathbf{P}_{0}^{2}}{2M}+\frac{\mathbf{P}^{2}}{2\mu}-\frac{1}{|\mathbf{r}|}+\mathbf{E}(t)\mathbf{r},\label{eq:quant_hamiltonian_main}
\end{equation}
which is separable in these coordinates, thus the solution can be
carried out in the two subsystems independently: 
\begin{equation}
\Psi_{ec}({\bf r}_{e},{\bf r}_{c},t)=\Psi({\bf r},t)\Psi_{0}({\bf r}_{0},t),\label{eq:ent_calc_psi_ec}
\end{equation}
where the coordinates of the two sides are connected via the transformation
(\ref{eq:quant_twobody_transform}). This very step, however, while
separates the problem in the coordinates chosen, does still involve
the entanglement of the individual particles in $\Psi_{ec}$.

\subsection{Subsystem: center of mass motion\label{sub:strong-field-center-of-mass}}

The center of mass part of the Hamiltonian describes a free-particle
propagation via the time-dependent Schrödinger equation (TDSE) 
\begin{equation}
i\frac{\partial}{\partial t}\Psi_{0}=H_{0}\Psi_{0}\text{ \thinspace\thinspace\thinspace\thinspace\thinspace\thinspace\ with \thinspace\thinspace\thinspace\thinspace\thinspace\thinspace}H_{0}=\frac{\mathbf{P}_{0}^{2}}{2M},\label{eq:quant_center_tdse}
\end{equation}
We assume that $\Psi_{0}$ is initially a localized Gaussian wave
packet at rest in coordinate space, which yields the solution of (\ref{eq:quant_center_tdse})
as 
\begin{equation}
\Psi_{0}(\mathbf{r}_{0},t)=\left(\frac{\sigma/\sqrt{\pi}}{\sigma^{2}+i\frac{t}{M}}\right)^{3/2}\exp\left(-\frac{\mathbf{r}_{0}^{2}}{2\left(\sigma^{2}+i\frac{t}{M}\right)}\right).\label{eq:quant_center_psi}
\end{equation}
We set the parameter $\sigma=1.$ i.e. a Bohr radius. This is the
well known free wave packet with root mean square deviations of the
center of mass coordinates in each direction spreading as 
\begin{equation}
\Delta x_{0}=\Delta y_{0}=\Delta z_{0}=\sqrt{\sigma^{2}+t^{2}/M^{2}\sigma^{2}}
\end{equation}
which is to be evaluated for the time interval given by the duration
$T_{\text{max}}$ of the exciting pulse. The typical value of the
latter in strong field experiments is $T_{\text{max}}=300$ a.u. corresponding
to a few femtoseconds, used also in our simulations. Due to the large
value of $M\approx1837$, the spreading during the interaction is
negligible: around 1.3\% of the original width, which will help us
to make the effect of the laser pulse on the quantum entropies more
visible in Section \ref{sec:results}.

\subsection{Subsystem: relative motion\label{sub:strong-field-relative}}

We assume a linearly polarized laser field which is the usual scenario
in many strong-field processes. \ This suggests to use cylindrical
coordinates $\rho=\sqrt{x^{2}+y^{2}}$, $\varphi$\ and $z$, the
latter being the direction of the external electric field strength.
We shall seek solutions that start from the ground state of the Coulomb
potential at $t=0$. This does not depend on the azimuthal angle $\varphi$
and this remains valid for the solution at any later time. Then the
wave function of the relative motion $\Psi\left(z,\rho,t\right)$
obeys the axially symmetric three-dimensional time-dependent Schrödinger
equation: 
\begin{equation}
i\frac{\partial}{\partial t}\Psi\left(z,\rho,t\right)=\left[T_{z}+T_{\rho}+V(z,\rho,t)\right]\Psi\left(z,\rho,t\right)\label{eq:quant_rel_tdse}
\end{equation}
where the two relevant terms of the kinetic energy operator are given
by 
\begin{equation}
T_{z}=-\frac{1}{2\mu}\frac{\partial^{2}}{\partial z^{2}},\,\,\,\, T_{\rho}=-\frac{1}{2\mu}\left[\frac{\partial^{2}}{\partial\rho^{2}}+\frac{1}{\rho}\frac{\partial}{\partial\rho}\right],\label{eq:quant_rel_kinetic}
\end{equation}
and $V$ includes both the Coulomb and the time-dependent external
potential: 
\begin{equation}
V(z,\rho,t)=-\frac{1}{\sqrt{z^{2}+\rho^{2}}}+z\cdot E_{z}(t).\label{eq:quant_rel_potential}
\end{equation}

Because we are working in the nonperturbative regime, an analytical
solution of (\ref{eq:quant_rel_tdse})-(\ref{eq:quant_rel_potential})
is not possible, so we have to resort to a numerical procedure. For
the efficient numerical solution of the above problem in real space,
we have developed a novel algorithm \citep{majorosi2016tdsesolve}
that we called hybrid splitting method, which is built on the combination
of the fourth order finite difference approximation in the 2D Crank-Nicolson
method and the (high order) split-operator methods. The main feature
of the algorithm is that it incorporates the Coulomb singularity and
the singularity of $T_{\rho}$ directly using the required Neumann
and Robin boundary conditions 
\begin{equation}
\lim_{\rho\rightarrow0}\left.\frac{\partial\Psi}{\partial\rho}\right\vert _{z\neq0}=0\text{,\thinspace\thinspace\thinspace\thinspace\thinspace\ and\thinspace\thinspace\thinspace\thinspace\thinspace\thinspace\thinspace}\left.\left[\frac{\partial}{\partial\rho}+\mu\right]\Psi\right\vert _{r=0}=0\label{eq:quant_rel_boundary}
\end{equation}
at the gridpoints on the symmetry axis ($\rho=0$). We can achieve
reasonable accuracy already at the uniform spatial discretization
step size $\Delta z=\Delta\rho=0.2,$ which may seem to be rough at
first sight, but it turns out to be sufficient \citep{majorosi2016tdsesolve}
in view of the large extension of the ionized part of the relative
wave function. For all the simulations presented in this paper, the
initial state is the 1s ground state with energy $\varepsilon_{0}=-\mu/2$
and $\mu=0.999456$, corresponding to the reduced mass of the proton-electron
system.

\subsection{Characterization of the ionization\label{sub:strong-field-dynamics}}

Now we discuss the properties of the expected ionization process and
some general features of the dynamics of the system. First, we assume
an external field of the form 
\begin{equation}
E_{z}(t)=Fg(t)\cos\left(\omega t\right),\label{eq:quant_Ez}
\end{equation}
where $F$ is the parameter of the amplitude of the external electric
field and $g(t)$ gives the finite pulse shape which is scaled so
that its minima are $0$ and its maxima are $1$. We will use the
particular form of (\ref{eq:sim_sinpulse3_E}) later in this article.
We assume $E_{z}(t)=0$ for $t\leq0$.

Regarding the electric field amplitude parameter $F$, there is a
specific value $F_{\mathrm{tu}}$ that separates two regimes, in which
the system has distinct behavior. In the \emph{tunneling ionization
}regime $F<F_{\mathrm{tu}}$ \ there is always a potential barrier
$V(F_{\mathrm{tu}},t)>\varepsilon_{0}$ in the vicinity of the atom,
while in the o\emph{ver-the-barrier ionization }regime $F>F_{\mathrm{tu}}$
this barrier does vanish to a varying extent both in space and time,
determined by $F$ and by the shape of the laser pulse. By solving
for $z=z\left(\varepsilon_{0},F_{\mathrm{tu}}\right)$ in 
\begin{equation}
\varepsilon_{0}=-\frac{1}{z}+z\cdot F_{\mathrm{tu}}\label{eq:quant_eqz_tu}
\end{equation}
at cross section $\rho=0$, a quick calculation reveals that this
critical value is $F_{\mathrm{tu}}$ $=$ $|\varepsilon_{0}|^{2}/4$,
i.e. $F_{\mathrm{tu}}\approx0.0624$.

Since the external field affects only the dynamics of the relative
core-electron motion, we can use certain physical quantities calculated
only from the relative wave function to describe its effects. From
the several possibilities we picked only two of them.

The first is the $z$ component of the mean velocity, i.e. the average
of the relative probability current density, given by 
\begin{equation}
\overline{v}_{z}(t)=\mathrm{Im}\left\langle \Psi(t)|\partial_{z}\Psi(t)\right\rangle /\mu.\label{eq:quant_vz_t}
\end{equation}
This gives information about the kinematical properties of the ``classical\textquotedblright \ particle
which behaves according to the Ehrenfest theorems. (We note that there
can be no mean displacement in the transverse directions $x$, $y$
because of the dipole approximation we use.)

The other descriptive time-dependent quantity we use, is based on
the projection onto the initial state 
\begin{equation}
f(t)=1-\left\vert \left\langle \Psi(0)|\Psi(t)\right\rangle \right\vert ^{2}\label{eq:quant_ion_project}
\end{equation}
which is actually the loss of the ground state population. This $f(t)$
can also be interpreted as the probability of leaving the vicinity
of the center of mass ($\mathbf{r}_{0}=0$). We have found that $f(T_{\text{max}})$
is a good indicator of the fraction of ionization, incorporated in
a continuum wave packet, because even the largest populations of the
excited bound hydrogen states turn out to be an order of magnitude
smaller than the ground state population loss in this strong field
process. This has been verified numerically\ in our actual calculations,
and this feature of (\ref{eq:quant_ion_project}) was also utilized
in other strong-field calculations like the well-known Lewenstein
model \citep{lewenstein1994hhgtheory}.

We will use these quantities for the analysis of the entanglement
dynamics, illustrating how much the atom was ionized and approximately
in which direction the particle moves.

\section{Entanglement calculation\label{sec:entanglement-calculation}}

According to the standard procedure of calculating the entanglement
(for more details see Appendix \ref{sec:quantum-entanglement}) we
need first the density matrix of the composite system 
\begin{equation}
\varrho_{ec}(\mathbf{r}_{e}^{\prime},\mathbf{r}_{e},\mathbf{r}_{c}^{\prime},\mathbf{r}_{c},t)=\Psi_{ec}^{\ast}(\mathbf{r}_{e}^{\prime},\mathbf{r}_{c}^{\prime},t)\Psi_{ec}(\mathbf{r}_{e},\mathbf{r}_{c},t)\label{eq:ent_dm_two_particle2}
\end{equation}
and then the reduced single particle density matrices that are obtained
by tracing over the other particle's degrees of freedom: 
\begin{equation}
\varrho_{c}(\mathbf{r}_{c}^{\prime},\mathbf{r}_{c},t)=\mathrm{Tr}_{e}\left[\hat{\varrho}_{ec}\right]=\int\varrho_{ec}(\mathbf{r}_{e},\mathbf{r}_{e},\mathbf{r}_{c}^{\prime},\mathbf{r}_{c},t){\rm d}\mathbf{r}_{e}^{3},\label{eq:ent_dm_core2}
\end{equation}
\begin{equation}
\varrho_{e}(\mathbf{r}_{e}^{\prime},\mathbf{r}_{e},t)=\mathrm{Tr}_{c}\left[\hat{\varrho}_{ec}\right]=\int\varrho_{ec}(\mathbf{r}_{e}^{\prime},\mathbf{r}_{e},\mathbf{r}_{c},\mathbf{r}_{c},t){\rm d}\mathbf{r}_{c}^{3}.\label{eq:ent_dm_electron2}
\end{equation}
As it is known, a good measure of bipartite entanglement is the Neumann
entropy:
\begin{equation}
S_{N}(t)=-\mbox{Tr}\left[\hat{\varrho}_{e}(t)\ln\hat{\varrho}_{e}(t)\right]=-\mbox{Tr}\left[\hat{\varrho}_{c}(t)\ln\hat{\varrho}_{c}(t)\right].\label{eq:ent_ent_neumann2}
\end{equation}
In our case $\Psi_{ec}$ of Eq. (\ref{eq:ent_calc_psi_ec}) contains
$\Psi_{0}$ as given analytically by Eq. (\ref{eq:quant_center_psi}),
while the relative part $\Psi$ is available only numerically in cylindrical
coordinates $(z;\rho)$, i.e it is a large two-dimensional array of
numbers.

\subsection{The necessity of a different approach\label{sub:calculation-approach}}

Now, if we try to apply the machinery of Eqs. (\ref{eq:ent_dm_two_particle2})-(\ref{eq:ent_ent_neumann2})
with the discrete function (\ref{eq:ent_calc_psi_ec}), we can quickly
conclude that the array size of the discretized density matrices involved
will be prohibitively high. If we try to perform the reduction (\ref{eq:ent_dm_core2})
and then to calculate (\ref{eq:ent_ent_neumann2}), we face effectively
an $\sim N^{9}$ operations count per one value of Neumann entropy,
where $N$ is the characteristic number of gridpoints of a spatial
coordinate. Using a typical setup, we needed at least $N=10^{3}$
gridpoints to contain the ionized electron waves. So if we make an
optimistic guess that the execution takes about 1 second per $10^{9}$
operations, then obtain a runtime of $32\times10^{9}$ years. This
makes the computation according to the standard approach of Appendix
\ref{sec:quantum-entanglement} practically impossible, thus we have
to find a viable approximation.

\subsection{Directionally reduced dynamics\label{sub:entanglement-reduced-dynamics}}

We propose to circumvent the prohibitively large numerical load of
the problem by restricting ourselves to only one coordinate direction
(at a time), i.e. to utilize the directionally reduced dynamics. Since
the system is axially symmetric around the polarization direction
of the laser field (the axis $\rho=0$ at all times), it seems to
be plausible to assume that the interesting physics happens in this
direction. However, we are also going to use the information contained
in the perpendicular directions.

The directionally reduced density matrices of the relative part are
the following 
\begin{equation}
\varrho_{x}(x^{\prime},x)=\iint\Psi^{\ast}(x^{\prime},y,z)\Psi(x,y,z)\text{d}z\text{d}y,\label{eq:ent_calc_dm_x}
\end{equation}
\begin{equation}
\varrho_{z}(z^{\prime},z)=2\pi\int\Psi^{\ast}(z^{\prime},\rho)\Psi(z,\rho)\rho\text{d}\rho\label{eq:ent_calc_dm_z}
\end{equation}
and because of the axial symmetry we have for the $y$ direction 
\begin{equation}
\varrho_{x}=\varrho_{y}.\label{eq:ent_calc_symm_axial}
\end{equation}
The directionally reduced density matrix of the center of mass part
is 
\begin{equation}
\varrho_{0,z}(z_{0}^{\prime},z_{0})=2\pi\int\Psi_{0}^{\ast}(z_{0}^{\prime},\rho_{0})\Psi_{0}(z_{0},\rho_{0})\rho_{0}\text{d}\rho_{0}.\label{eq:ent_calc_dm_z_0}
\end{equation}
Due to the assumed Gaussian form (\ref{eq:quant_center_psi}) it is
a pure state density matrix which can be calculated analytically.
Again, because of symmetry we have in the other directions 
\begin{equation}
\varrho_{0,z}=\varrho_{0,y}=\varrho_{0,x}.\label{eq:ent_calc_symm_gauss}
\end{equation}
In addition to this, only the density matrix of the relative part
must be evaluated numerically.

After we have completed these, we will utilize that the separability
is true in each direction: 
\begin{equation}
\varrho_{ec,x}=\varrho_{x}\otimes\varrho_{0,x}\text{ and }\varrho_{ec,z}=\varrho_{z}\otimes\varrho_{0,z}.\label{eq:ent_calc_dm_z_ec}
\end{equation}
Finally, we apply the necessary coordinate transformation (\ref{eq:quant_twobody_transform})
to (\ref{eq:ent_calc_dm_z_ec}), then the $x$ and $z$ directional
two particle reduced density matrices are given by 
\begin{equation}
\begin{aligned}\varrho_{ec,x}(x_{e}^{\prime},x_{e},x_{c}^{\prime},x_{c})=\varrho_{x}(x_{e}^{\prime}-x_{c}^{\prime},x_{e}-x_{c})\times\\
\varrho_{0,x}(\alpha_{e}x_{e}^{\prime}+\alpha_{c}x_{c}^{\prime},\alpha_{e}x_{e}+\alpha_{c}x_{c})
\end{aligned}
\label{eq:ent_calc_dm_x_ec_tran}
\end{equation}
\begin{equation}
\begin{aligned}\varrho_{ec,z}(z_{e}^{\prime},z_{e},z_{c}^{\prime},z_{c})=\varrho_{z}(z_{e}^{\prime}-z_{c}^{\prime},z_{e}-z_{c})\times\\
\varrho_{0,z}(\alpha_{e}z_{e}^{\prime}+\alpha_{c}z_{c}^{\prime},\alpha_{e}z_{e}+\alpha_{c}z_{c})
\end{aligned}
\label{eq:ent_calc_dm_z_ec_tran}
\end{equation}
Then, we calculate the subsystem density matrices as in Appendix \ref{sec:quantum-entanglement}.
Therefore, the one dimensional reduced density matrices of the core
coordinates are 
\begin{equation}
\varrho_{c,x}(x_{c}^{\prime},x_{c})=\text{Tr}_{e}\left[\hat{\varrho}_{ec,x}\right]=\int\varrho_{ec,x}(x_{e},x_{e},x_{c}^{\prime},x_{c})\text{d}x_{e},\label{eq:ent_calc_dm_x_c}
\end{equation}
\begin{equation}
\varrho_{c,z}(z_{c}^{\prime},z_{c})=\text{Tr}_{e}\left[\hat{\varrho}_{ec,z}\right]=\int\varrho_{ec,z}(z_{e},z_{e},z_{c}^{\prime},z_{c})\text{d}z_{e},\label{eq:ent_calc_dm_z_c}
\end{equation}
and, similarly, we have for the electron coordinates 
\begin{equation}
\varrho_{e,x}(x_{e}^{\prime},x_{e})=\text{Tr}_{c}\left[\hat{\varrho}_{ec,x}\right]=\int\varrho_{ec,x}(x_{e}^{\prime},x_{e},x_{c},x_{c})\text{d}x_{c}.\label{eq:ent_calc_dm_x_e}
\end{equation}
\begin{equation}
\varrho_{e,z}(z_{e}^{\prime},z_{e})=\text{Tr}_{c}\left[\hat{\varrho}_{ec,z}\right]=\int\varrho_{ec,z}(z_{e}^{\prime},z_{e},z_{c},z_{c})\text{d}z_{c}.\label{eq:ent_calc_dm_z_e}
\end{equation}
In this way we have the building blocks of the two-body Coulomb system
as six pieces of one dimensional reduced density matrices, which can
already be computed in a reasonable amount of time.

\subsection{Correlation quantification per direction\label{sub:entanglement-directional-entropy}}

From these reduced density matrices we can calculate several quantum
entropies, and each has a specific interesting aspect, we will list
them in the following. For simplicity, we use mainly formulae of the
von Neumann entropy, and we usually drop its subscript N.

\emph{Spatial entropy}: this can be calculated from the reduced density
matrix of the relative part as 
\begin{equation}
S_{z}(t)=S_{N}(\varrho_{z}(t))=-\sum\limits _{k}\lambda_{k}^{(z)}(t)\ln\lambda_{k}^{(z)}(t),\label{eq:ent_SN_dmz}
\end{equation}
where $\lambda_{k}^{(z)}(t)$ are the eigenvalues of $\varrho_{z}(t)$.
We shall call (\ref{eq:ent_SN_dmz}) ``spatial entanglement\textquotedblright \ measure,
because it quantifies the entanglement between the coordinates $z$,
$\rho$ (or the nonseparability of the numerical solution) according
to the theory of pure bipartite systems. It is also the entropy of
the two dimensional subspace 
\begin{equation}
S_{ec,z}(t)=S_{N}(\varrho_{z}(t)\otimes\varrho_{0,z}(t))=S_{z}(t),\label{eq:ent_SN_sep}
\end{equation}
since $S_{N}(\varrho_{0,z}(t))=0$. We also note that using 
\begin{equation}
S_{x}(t)=S_{N}(\varrho_{x}(t))=-\sum\limits _{k}\lambda_{k}^{(x)}(t)\ln\lambda_{k}^{(x)}(t),\label{eq:ent_SN_dmx}
\end{equation}
where $\lambda_{k}^{(x)}(t)$ are the eigenvalues of $\varrho_{x}(t)$,
is also an option as a ``spatial entanglement\textquotedblright \ measure.
However, since the laser polarization coincides with the $z$ axis,
it is the most interesting to know the nonseparability between the
$z$ and $x\otimes y$ subspaces, therefore, we will prefer the use
of $S_{z}(t)$.

\emph{Average mutual entropy per direction}: as introduced by the
formula (\ref{eq:qinf_ent_mut_entang}), the (average) quantum mutual
entropy is a true nonseparability and correlation measure generally,
which can be used between the single coordinate subsystems of the
electron and the ion-core in a given direction. They are written along
the $x$ and $z$ direction as 
\begin{equation}
\overline{S}(x_{e}:x_{c},t)=\frac{1}{2}\left[S_{e,x}(t)+S_{c,x}(t)-S_{x}(t)\right],\label{eq:ent_SN_mut_avg_x}
\end{equation}
\begin{equation}
\overline{S}(z_{e}:z_{c},t)=\frac{1}{2}\left[S_{e,z}(t)+S_{c,z}(t)-S_{z}(t)\right].\label{eq:ent_SN_mut_avg_z}
\end{equation}
To remind, these are exact formulae for pure bipartite states. However,
these measures combine classical and entanglement related correlations
otherwise, and in order to apply them as entanglement measures (per
direction), we need to look at all of their constituent parts. It
is also interesting how the conditional entropies behave.

\emph{Core entropies per direction}: as we will show below, the following
quantum entropies 
\begin{equation}
S_{c,x}(t)=S_{N}(\varrho_{c,x}(t))=-\sum\limits _{k}\lambda_{k}^{(c,x)}(t)\ln\lambda_{k}^{(c,x)}(t),\label{eq:ent_SN_dmx_c}
\end{equation}
\begin{equation}
S_{c,z}(t)=S_{N}(\varrho_{c,z}(t))=-\sum\limits _{k}\lambda_{k}^{(c,z)}(t)\ln\lambda_{k}^{(c,z)}(t),\label{eq:ent_SN_dmz_c}
\end{equation}
where $\lambda_{k}^{(c,x)}(t)$, $\lambda_{k}^{(c,z)}(t)$ are the
eigenvalues of $\varrho_{c,x}(t)$ and $\varrho_{c,z}(t)$ respectively,
measure approximately the particle-particle correlation direction-wise.
The reason is the following: because of the orders of magnitude of
mass difference present in the coordinate transformation (\ref{eq:ent_calc_dm_z_ec_tran}),
the reduced density matrix $\varrho_{c,z}$ will be close to $\varrho_{0,z}$.
This causes that only a tiny fraction ($m_{e}/M$) of the entropy
$S_{z}(t)$ of $z_{e}\otimes z_{c}$ is transferred to subsystem $z_{c}$,
because the mass difference suppresses the eigenvalues and eigenvectors
of $\varrho_{z}$. Knowing that $\varrho_{0,z}$ is a pure state density
matrix with zero entropy, we conclude that additional surplus values
in entropy $S_{c,z}(t)$ quantifies a particle-particle correlation
along the $z$ direction. In other words, it is the nonseparability
between $z_{c}$ and $z_{e}$, which yet to be called entanglement.
The same considerations also apply to the $x$ direction. Because
these Neumann entropies are actually correlation entropies in this
case, we expect them look really similar to the respective quantum
mutual entropies. For sake of completeness, we note that the entropies
(\ref{eq:ent_SN_dmx_c}) and (\ref{eq:ent_SN_dmz_c}) are also entanglement
entropies of two special bipartitions of the six coordinate quantum
system, namely $x_{c}$ against all the other coordinates and $z_{c}$
against all the other coordinates, respectively.

\emph{Electron entropies per direction}: these are also of importance
related to the conditional entropies, and the distinction of quantum
versus classical correlations. Similarly, they are also special entanglement
entropies of two bipartitions of the six coordinate quantum system
in similar manner as the core entropies per direction. They are calculated
as 
\begin{equation}
S_{e,x}(t)=S_{N}(\varrho_{e,x}(t))=-\sum\limits _{k}\lambda_{k}^{(e,x)}(t)\ln\lambda_{k}^{(e,x)}(t),\label{eq:ent_SN_dmx_e}
\end{equation}
\begin{equation}
S_{e,z}(t)=S_{N}(\varrho_{e,z}(t))=-\sum\limits _{k}\lambda_{k}^{(e,z)}(t)\ln\lambda_{k}^{(e,z)}(t),\label{eq:ent_SN_dmz_e}
\end{equation}
where $\lambda_{k}^{(e,x)}(t)$, $\lambda_{k}^{(e,z)}(t)$ are the
eigenvalues of $\varrho_{e,x}(t)$ and $\varrho_{e,z}(t)$ respectively.
We note that although $\varrho_{e}$ and $\varrho_{c}$ must have
the same eigenvalues, this won't be true for the reduced density matrices
$\varrho_{e,z}$ and $\varrho_{c,z}$ in direction $z$ if the values
of $S_{z}(t)$ are not negligible. (Same goes for the $x$ direction.)
Then the coordinate transformation (\ref{eq:ent_calc_dm_z_ec_tran})
causes that the major fraction ($m_{c}/M$) of the entropy $S_{z}(t)$
of $z_{e}\otimes z_{c}$ is transferred to subsystem $z_{e}$, because
the reduced density matrix $\varrho_{e,z}$ will be close to $\varrho_{z}$.
Based on the quantum information theoretic properties of the Neumann
entropies, this spurious eigenvalue contribution can be extracted,
but not completely. This ``eigenvalue extraction\textquotedblright \ we
refer to can be realized by the $S_{e,z}(t)-S_{z}(t)$ entropy subtraction,
these are the single direction \textit{negative quantum conditional
entropies} of the core $x$ and core $z$ reduced density matrices:
\begin{equation}
-S(x_{c}|x_{e},t)=S_{e,x}(t)-S_{x}(t),\label{eq:ent_SN_cond_xc}
\end{equation}
\begin{equation}
-S(z_{c}|z_{e},t)=S_{e,z}(t)-S_{z}(t).\label{eq:ent_SN_cond_zc}
\end{equation}
Since they are related to the correlation one way or the other, from
the above reasoning it follows that $-S(x_{c}|x_{e},t)$ and $-S(z_{c}|z_{e},t)$
should be similar to $S_{c,x}(t)$ and $S_{c,z}(t)$ and therefore
also to their mutual entropy, respectively. Based on this reasoning
we will see that the subsystems $x_{e}\otimes x_{c}$ and $z_{e}\otimes z_{c}$
are mainly subject to quantum entanglement (in accordance with (\ref{eq:qinf_ent_cond_quant})),
not classical correlation (also present), which we show in Section
\ref{sec:results}.

\textit{Upper bound of the core entropy}: using the strong subadditivity
of the Neumann entropy, an upper bound can be given for the true 3D
electron-core entanglement as 
\begin{equation}
S_{e}(t)=S_{c}(t)\leq S_{c,z}(t)+2S_{c,x}(t)=S_{\mathrm{bound}}(t),\label{eq:ent_SN_ec_bound}
\end{equation}
where the one dimensional core entropies were substituted into (\ref{eq:qinf_SN_subadd}),
because they tend to be smaller than those of the electrons and because
of the physical reasons outlined above. The equation (\ref{eq:ent_SN_ec_bound})
serves also as a good analytical criteria that we should fulfill with
an approximate formula for electron-core entanglement.

\subsection{Approximation of the entanglement\label{sub:entanglement-approximation}}

Now we introduce our approximate entanglement measure, which is one
of the main purposes of this paper. 

We approximate the pure state of our six-dimensional quantum system
by replacing it with 
\begin{equation}
\varrho_{ec}^{(\mathrm{sep})}(t)=\varrho_{ec,x}(t)\otimes\varrho_{ec,y}(t)\otimes\varrho_{ec,z}(t),\label{eq:ent_Sec_dm_sep}
\end{equation}
which is separable direction-wise but it includes the $\varrho_{ec,x}(x_{e}^{^{\prime}},x_{e},x_{c}^{^{\prime}},x_{c})$,
$\varrho_{ec,y}(y_{e}^{^{\prime}},y_{e},y_{c}^{^{\prime}},y_{c})$,
$\varrho_{ec,z}(z_{e}^{^{\prime}},z_{e},z_{c}^{^{\prime}},z_{c})$
two dimensional reduced density matrices, which contain all the pair
correlations between the coordinates $x_{e}$-$x_{c}$, $y_{e}$-$y_{c}$,
$z_{e}$-$z_{c}$, respectively. (Because of the symmetry, the physics
in the subspaces $x$ and $y$ are identical, so (\ref{eq:ent_calc_symm_axial})
is true.) Then we obtain the entropy of $\varrho_{ec}^{(\mathrm{sep)}}$
from the additivity of the Neumann entropy (valid for separable systems),
and using that $S_{ec,j}(t)=S_{j}(t)+S_{0,j}(t)$, $\varrho_{ec,j}=\varrho_{j}\otimes\varrho_{0,j}$,
$j=x,y,z$ as 
\begin{equation}
S_{ec}^{(\mathrm{sep})}(t)=S_{x}(t)+S_{y}(t)+S_{z}(t),\label{eq:ent_Sec_Sec}
\end{equation}
The single-particle core and electron reduced density matrices read
\begin{equation}
\varrho_{e}^{(\mathrm{sep})}(t)=\varrho_{e,x}(t)\otimes\varrho_{e,y}(t)\otimes\varrho_{e,z}(t),\label{eq:ent_Sec_dm_e}
\end{equation}
\begin{equation}
\varrho_{c}^{(\mathrm{sep})}(t)=\varrho_{c,x}(t)\otimes\varrho_{c,y}(t)\otimes\varrho_{c,z}(t),\label{eq:ent_Sec_dm_c}
\end{equation}
with the standard definitions ($j=x,y,z$): 
\begin{equation}
\varrho_{e,j}(t)=\mathrm{Tr}_{c}\left[\varrho_{ec,j}(t)\right]\text{ and }\varrho_{c,j}(t)=\mathrm{Tr}_{e}\left[\varrho_{ec,j}(t)\right].\label{eq:ent_Sec_dm_j_e_c}
\end{equation}
For the entropies of these, the following hold: 
\begin{equation}
S_{e}(t)=S_{e,x}(t)+S_{e,y}(t)+S_{e,z}(t),\label{eq:ent_Sec_S_e}
\end{equation}
\begin{equation}
S_{c}(t)=S_{c,x}(t)+S_{c,y}(t)+S_{c,z}(t).\label{eq:ent_Sec_S_c}
\end{equation}
We propose to quantify the total entanglement between $e$ and $c$
based on the average mutual entropy (\ref{eq:qinf_ent_mut_entang})
as 
\begin{equation}
\overline{S}_{ec}(e:c,t)=\frac{1}{2}S(e:c,t)=\frac{1}{2}\left[S_{e}(t)+S_{c}(t)-S_{ec}(t)\right].\label{eq:ent_Sec_avg}
\end{equation}
After rearranging the terms and using symmetry relations (\ref{eq:ent_calc_symm_axial}),
(\ref{eq:ent_calc_symm_gauss}) we obtain 
\begin{equation}
\overline{S}_{ec}(e:c,t)=\frac{1}{2}\left[2S(x_{e}:x_{c},t)+S(z_{e}:z_{c},t)\right]\label{eq:ent_Sec_mut_avg_formula}
\end{equation}
as the final form of our approximate formula for the total entanglement.

The introduction of the factor $1/2$ in the above definition is useful
in the case when each of the two dimensional subsystems are in a pure
state, i.e. $S_{ec,x}(t)=S_{ec,y}(t)=S_{ec,z}(t)=0$. Then, the bipartite
Schmidt theorem holds in these subspaces as $S_{e,j}(t)=S_{c,j}(t)$
(with $j=x,y,z$), and we obtain 
\begin{equation}
\overline{S}_{ec}^{(\mathrm{pure,sep)}}(t)=S_{c,x}(t)+S_{c,y}(t)+S_{c,z}(t),\label{eq:ent_Sec_mut_avg_pure}
\end{equation}
which is by definition the exact entanglement measure.

\subsection{Connection to a one-dimensional approximation\label{sub:entanglement-one-dimensional}}

To ensure the comparability of our new results with our earlier one-dimensional
simulations \citep{czirjak2013rescatterentanglement}, we briefly
discuss those now in relation to the previous section. Let us assume
that the potential of the relative coordinate quantum system can be
approximated in as 
\begin{equation}
V(x,y,z,t)=V_{z}(z,t)+V_{x}(x)+V_{y}(y),\label{eq:separable_potential}
\end{equation}
i.e. the 3D Coulomb potential is replaced by some one dimensional
model potentials and the electric dipole term is contained in $V_{z}(z,t)$
only. Certain simple 3D models can be treated with this approach also
analytically: an example of these is the Moshinsky atom \citep{bouvrie2014entmoshinsky,march2008moshinskycorrel}
with a single electron in an external laser field using dipole approximation.

Starting the numerical simulations with a potential of the form (\ref{eq:separable_potential})
from the separable ground state of the relative system, the system
will stay separable along the $x,\, y,\, z$ directions with the relative
wave function of form 
\begin{equation}
\Psi(x,y,z,t)=\psi_{x}(x)\psi_{y}(y)\psi_{z}(z,t)e^{-i(\varepsilon_{0,x}+\varepsilon_{0,y})t},\label{eq:ent_Sec_one_state}
\end{equation}
thus its density matrix will have the time-dependent form: 
\begin{equation}
\varrho(t)=\varrho_{x}\otimes\varrho_{y}\otimes\varrho_{z}(t).\label{eq:ent_Sec_one_dm_rel}
\end{equation}
From (\ref{eq:ent_Sec_one_dm_rel}) and (\ref{eq:ent_calc_dm_z_ec})
it follows that (\ref{eq:ent_Sec_avg}) is an exact entanglement measure,
as it yields (\ref{eq:ent_Sec_mut_avg_pure}) as mentioned before.
Then the complete entanglement dynamics induced by the laser field
is restricted to the $z_{c}\otimes z_{e}$ subspace, described by
$S_{c,z}(t)$. Due to the large mass ratio in (\ref{eq:quant_center_psi}),
the entanglement of the perpendicular directions changes order of
magnitudes slower than $S_{c,z}(t)$ does, thus it can be regarded
as a constant shift.

\section{Results\label{sec:results}}

\subsection{External electric field\label{sub:results_field}}

In our simulations, we expose the hydrogen atom to a few cycle laser
pulse with a sin-squared envelope function. The corresponding time-dependent
electric field has nonzero values only in the interval $0\leq t\leq3T$
according to the formula: 
\begin{equation}
E_{z}(t)=F\cdot\sin^{2}\left(\frac{\pi t}{6T}\right)\cos\left(\frac{2\pi t}{T}+\mathrm{CEP}\cdot\pi\right),\label{eq:sim_sinpulse3_E}
\end{equation}
where $T$ is the period of carrier wave, $F$ is the strength of
the electric field, $\mathrm{CEP}\cdot\pi$ is the carrier envelope
phase. We keep the wavelength of the laser field through parameter
$T$ the same in all of the simulations: we set $T=100$ which corresponds
to a $\sim725\mathrm{nm}$ near infrared carrier wave. Varying the
parameter $F$ and separately the parameter $\mathrm{CEP}$, we investigate
the dynamics of the system with the emphasis on quantum entanglement.

\subsection{Simulation procedure\label{sub:results_procedure}}

The simulation of the time evolution starts from the ground state
of the relative Hamiltonian which was found by imaginary time propagation
having the energy $\varepsilon_{0}\approx-0.49972$. Other parameters
used in the numerical simulations of the relative wave function are
$(i)$ discretization parameters $\Delta z=\Delta\rho=0.2$, and $\Delta t=0.01$
for the fourth order splitting formula of \citep{majorosi2016tdsesolve},
$(ii)$ the total simulated time is $330$ atomic time units, $(iii)$
absorbing imaginary potentials are not used, $(iv)$ the simulation
box size is varied with parameter $F$. The dimensions of the latter
are $z_{\min}=-500$, $z_{\max}=500$, $\rho_{\max}=300$ for $F=0.1$.
For the evaluation of the partial derivatives $z$,$\rho$ we use
fourth order finite differences, and for the evaluation of the integrals
we also use a discrete sum approximation, both of which can be found
in \citep{majorosi2016tdsesolve}. After this, we perform the reduced
density matrix based calculations at each atomic time unit.

\subsection{Dynamics\label{sub:results_dynamics}}

We begin our analysis discussing the time-dependence of the ground
state population loss (\ref{eq:quant_ion_project}) of the relative
wave function, which is shown in Fig. \ref{fig:sinpulse3_F_projection}
for several values of $F$. We can see sudden increases of the ground
state population loss that are happening at the local extrema of the
electric field, more and more clearly as $F$ increases. As we have
already discussed in Section \ref{sec:strong-field}, we make distinction
between the tunneling ionization regime and the over-the-barrier ionization
regime, regarding the dynamics dependent on $F$: starting from the
former, we see from Fig. \ref{fig:sinpulse3_F_projection} that even
for $F=0.06$, (just below the over-the-barrier threshold) the total
ground state population loss is small ($0.02$) which implies small
amount of ionization in the tunneling regime. At $F=0.10$ we have
already a significant total ground state population loss ($0.33$)
with prominent over-the-barrier ionization. At the highest $F$ shown
($F=0.12$), the electric field increasingly dominates the Coulomb
force, and it almost doubles the total ground state population loss
($0.61$) and ionization.

\begin{figure}[H]
\includegraphics[width=1\columnwidth]{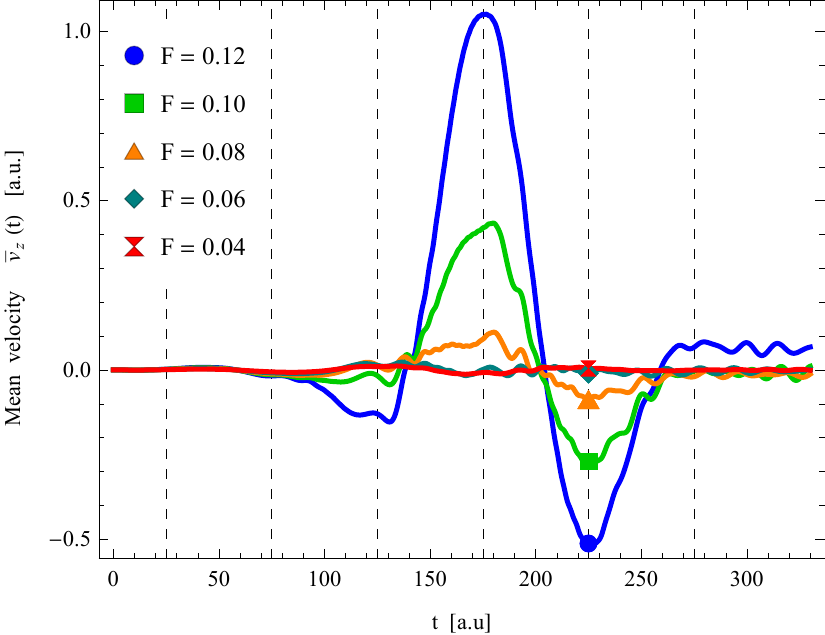}\protect\protect\caption{Time-dependence of the mean velocity of the relative wave function
$\overline{v}_{z}(t)$, defined in Eq. \eqref{eq:quant_vz_t}, for
the indicated values of the parameter $F$, with $\mathrm{CEP}=0$.
For $F=0.12$, this velocity somewhat correlates with the quiver motion
of the classical free electron moving under the influence of the same
uniform dipole electric field \eqref{eq:sim_sinpulse3_E}. The vertical
dashed lines denote the zero crossings of the electric field. (They
have the same meaning on all of the figures.) }

\label{fig:sinpulse3_F_velocity_z} 
\end{figure}

\begin{figure}[H]
\includegraphics[width=1\columnwidth]{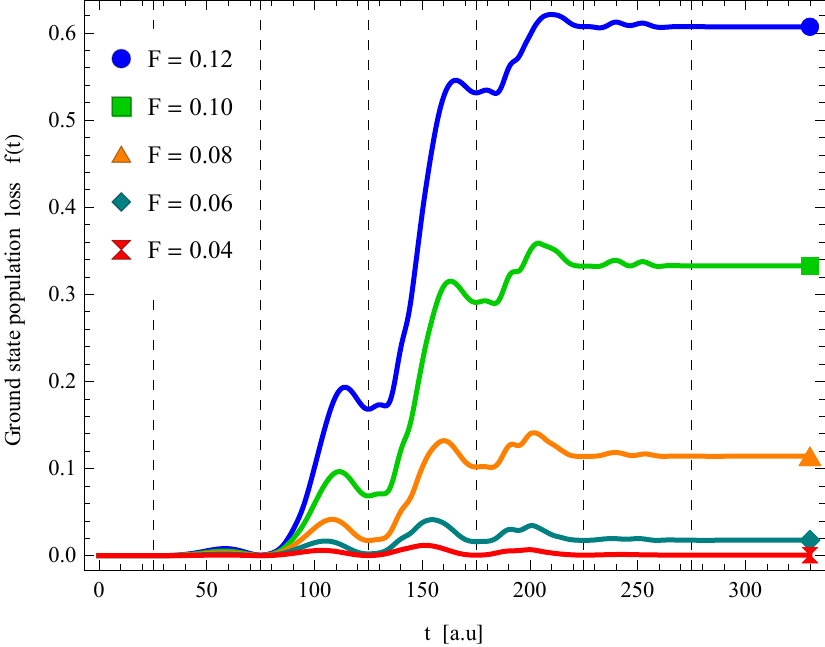}\protect\protect\caption{Time-dependence of the ground state population loss $f(t)$, defined
in Eq. \eqref{eq:quant_ion_project}, for the indicated values of
the parameter $F$, with $\mathrm{CEP}=0$. The $f(t)$ is linked
to the probability of ionization. For pure tunnel ionization i.e.
for $F<0.0624$, the ionization is very small. For higher values of
$F$, the $f(t)$ rises suddenly around the peaks of the laser pulse. }

\label{fig:sinpulse3_F_projection} 
\end{figure}

From Fig. \ref{fig:sinpulse3_F_velocity_z}, we can inspect how the
results translate to an averaged classical motion, using the mean
velocity component $\overline{v}_{z}(t)$ from the formula (\ref{eq:quant_vz_t}).
In the tunneling range (with $F=0.06$ or below), $\overline{v}_{z}(t)$
only slightly changes with time and has oscillating component, which
implies that the relative wave function is oscillating near the origin.
For amplitudes sufficiently above the over-the-barrier ionization
threshold $(F=0.1)$, the velocity somewhat correlates with the quiver
motion of the classical free electron moving under the influence of
the oscillating electric field (\ref{eq:sim_sinpulse3_E}). For example,
$\overline{v}_{z}(t)$ has local extrema near the zero crossings of
the electric field like within this classical picture. With increasing
$F$, the correlation of $\overline{v}_{z}(t)$ and this ``free\textquotedblright \ classical
motion becomes more clear, signaling the increase of importance of
the ionized waves. After the laser pulse ends, the ground state population
loss stops as expected, and $\overline{v}_{z}(t)$ appears to oscillate
near a constant mean value which is more remarkable with higher $F$.
(This latter value can be nonzero, which contradicts the mentioned
classical picture and the three step model.)

\subsection{Time-dependence of the quantum entropies\label{sub:results_entropies}}

Now we start to analyze the time-dependent dynamics of the quantum
entanglement of these ionization processes.

We begin the discussion of the various quantum entropies in the direction
parallel to the laser polarization axis ($z$), then in the direction
transverse to this polarization axis ($x$ or $y$) and in the last
paragraphs in this subsection, we conclude with the discussion of
the total electron-core entanglement approximated by our method. For
this task, we set the electric field parameter to be $F=0.1$ which
means an intermediate, over-the-barrier ionization range, and we choose
the carrier envelope phase to be $\mathrm{CEP}=0$.

\begin{figure}[H]
\includegraphics[width=1\columnwidth]{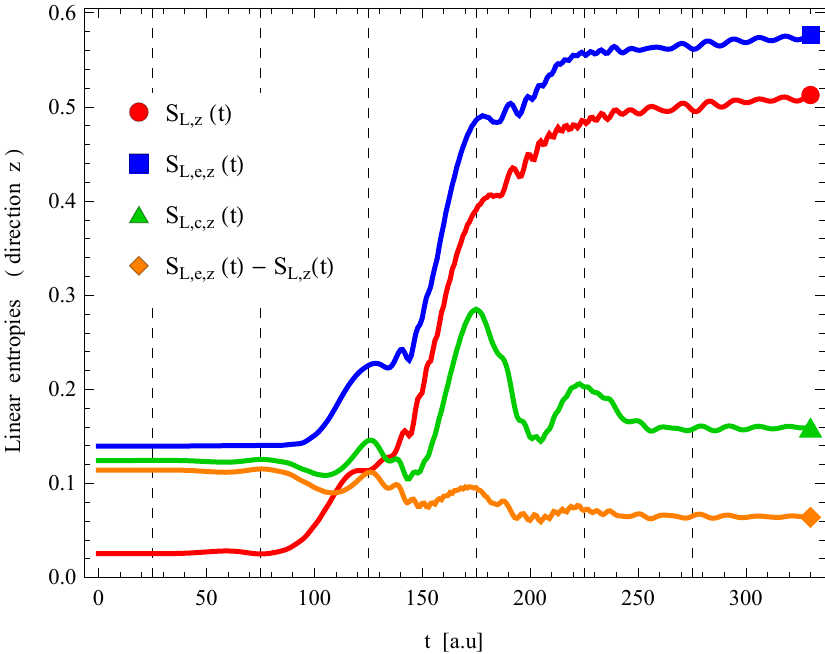}\protect\protect\caption{Comparison of the time-dependence of various linear entropies in direction
$z$, based on our reduced density matrix formalism, using $F=0.1$,
$\mathrm{CEP}=0$. Although the linear entropies of $S_{L,z}(t)$,
$S_{L,c,z}(t)$, $S_{L,e,z}(t)$ compare fairly well to the respective
Neumann entropies in Fig. \ref{fig:sinpulse3_F0.1_NeumEnt_Z}, the
negative linear conditional entropy $S_{L,e,z}(t)-S_{L,z}(t)$ gives
a false prediction. }

\label{fig:sinpulse3_F0.1_LinEnt_Z} 
\end{figure}

\begin{figure}[H]
\includegraphics[width=1\columnwidth]{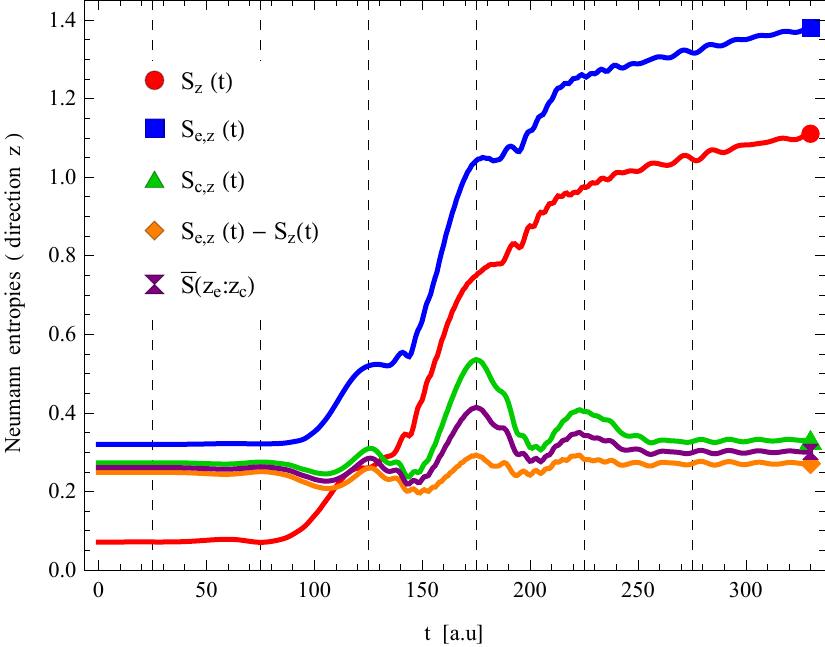}\protect\protect\caption{Comparison of the time-dependence of various Neumann entropies in
direction $z$, based on our reduced density matrix formalism, using
$F=0.1$, $\mathrm{CEP}=0$. The synchronous changes in $S_{c,z}(t)$,
in $S_{e,z}(t)-S_{z}(t)$, and in $\overline{S}(z_{e}:z_{c},t)$ signal
that they are related to a common source of correlation, which is
primarily the quantum entanglement between $z_{e}$ and $z_{c}$,
as evidenced by the high value of the negative conditional entropy
$S_{e,z}(t)-S_{z}(t)$ .}

\label{fig:sinpulse3_F0.1_NeumEnt_Z} 
\end{figure}

First, we discuss the linear entropies of the reduced density matrices
$\varrho_{z}$, $\varrho_{c,z}$ and $\varrho_{e,z}$. We use the
notation $S_{L,z}(t)$, $S_{L,c,z}(t)$, $S_{L,e,z}(t)$ and we plot
them in Fig. \ref{fig:sinpulse3_F0.1_LinEnt_Z} to provide them as
a comparison to the Neumann entropies $S_{z}(t)$, $S_{c,z}(t)$,
$S_{e,z}(t)$, which are shown in Fig. \ref{fig:sinpulse3_F0.1_NeumEnt_Z},
with the same parameters. Although these linear entropies compare
fairly well to the respective Neumann entropies, the orange line in
Fig. \ref{fig:sinpulse3_F0.1_LinEnt_Z} shows that the quantity $S_{L,e,z}(t)-S_{L,z}(t)$
gives false prediction, therefore we use only the Neumann entropies,
as we have already stated earlier.

The time-dependence of Neumann entropies corresponding to the direction
$z$ are shown in Fig. \ref{fig:sinpulse3_F0.1_NeumEnt_Z}. We see
that $S_{z}(t)$ and $S_{e,z}(t)$ share the main features but $S_{c,z}(t)$
has a different behavior. First, let us say some words about the time
dependence of the Neumann entropy $S_{z}(t)$ (plotted with red line).
Overall, this spatial entropy of the ``spatial entanglement'' between
$z$ and $\rho$ has major increase during the process due to the
ionization: it starts from a rather small value of $0.07$ and and
has a large permanent increase during the process (to the value $1.11$).
This entropy also continues to grow slowly but steadily after the
laser pulse ended, i.e. due to the mixing effect of the Coulomb potential
only. It has sudden increases in time near the peaks of the laser
pulse, however, with about 10 atomic time units of delays, with the
biggest jump occurring near the central peak. If we compare this plot
with the correspond curve of Fig. \ref{fig:sinpulse3_F_velocity_z}
we clearly see that the timings of these increases synchronize with
the increases of $|\bar{v}_{z}(t)|$. Regarding the small starting
value of $S_{z}(t)$, the initial state of $\exp(-\mu r)$ is almost
separable in $z$ and $\rho$, in accordance with the low value of
this entropy: particularly, the dominating eigenvalue of $\varrho_{z}$
at $t=0$ is $\lambda_{1}^{(z)}=0.9872$. This clearly shows that
the Neumann entropies, in general, take into account the other much
smaller eigenvalues in a more pronounced way.

In Fig. \ref{fig:sinpulse3_F0.1_NeumEnt_Z} we show the Neumann entropies
$S_{e,z}(t)$ and $S_{c,z}(t)$, and the negative conditional entropy
$S_{e,z}(t)-S_{z}(t)$, with blue, green and orange lines, respectively,
in order to attempt to answer the question, how the core-electron
correlation works in the directional $z$ subsystem. We stated in
Section \ref{sec:entanglement-calculation}, that the coordinate transformation
(\ref{eq:ent_calc_dm_z_ec_tran}) creates a special type of correlation.
Therefore, we should be able to acquire (at least partially) the correlation
information contained in $S_{c,z}(t)$ from the Neumann entropy $S_{e,z}(t)$.
As it is clearly shown by Fig. \ref{fig:sinpulse3_F0.1_NeumEnt_Z},
the majority of the time-dependent features of $S_{e,z}(t)$ seem
to be inherited from the Neumann entropy of the reduced density matrix
$\varrho_{z}(t)$ (for example, the sudden increases related to the
ionization), they are only shifted to higher values. However, if we
carefully inspect the curve of $S_{e,z}(t)-S_{z}(t)$ in Fig. \ref{fig:sinpulse3_F0.1_NeumEnt_Z}
(in orange) we can easily observe that its main features (like its
correlation with the laser pulse) are very similar to those of $S_{c,z}(t)$.
Because these quantities are close to each other, it means that in
this subsystem the major correlation is quantum entanglement, as we
stated earlier. Therefore, $\overline{S}(z_{c}:z_{e},t)$, defined
in (\ref{eq:qinf_ent_mut_entang}), can be used as an approximate
entanglement measure. We also make the observation that $S_{z}(t)$
is always upper bounded by $S_{e,z}(t)$, and the respective $z$
coordinate of the lighter electron contains more entropy than that
of the heavier ion-core, as expected.

Next, we discuss the time dependence of the resulting mutual entropy
$\overline{S}(z_{c}:z_{e},t)$, which is plotted in Fig. \ref{fig:sinpulse3_F0.1_NeumEnt_Z}
as a purple curve. This quantity inherits its features from $S_{c,z}(t)$
and $S_{e,z}(t)-S_{z}(t)$ by construction: it starts from an intermediate
value ($0.23$), rises and falls several times during the process,
contrary to $S_{z}(t)$. It stays almost constant after the laser
pulse, around a value ($0.25$) that is only slightly higher than
the initial value. The time-dependence of $\overline{S}(z_{c}:z_{e},t)$
correlates better with the shape of the laser pulse, and also has
much smaller peak value in the time window, than the aforementioned
spatial entropy. Interestingly, the rapid changes in ionization probability
during the process are not reflected by this particle-particle entanglement
of the $z$ directional subspace. The changes of this mutual entropy
are more correlated with the average velocity $\overline{v}_{z}(t)$,
which we expand more in the next subsection.

The curves of Fig. \ref{fig:sinpulse3_F0.1_NeumEnt_Z} clearly show
that the classical correlations also change under the effect of the
laser pulse: the gap between $S_{c,z}(t)$ and $S_{e,z}(t)-S_{z}(t)$
is dynamically increasing and decreasing, synchronously with the electric
field. Even though the respective mutual entropy includes these classical
effects, the also synchronous changes in $S_{c,z}(t)$ and in $S_{e,z}(t)-S_{z}(t)$
signal that the quantum entanglement behaves the same way, and the
high value of the negative conditional entropy causes it to be the
major correlation.

Here we ought to note that the actual related values of $S_{c,z}(t)$,
$\overline{S}(z_{c}:z_{e},t)$, $S_{e,z}(t)-S_{z}(t)$ are also influenced
by $\Psi_{0}$, that is by the adjustable parameter $\sigma^{2}$.
According to our simulations, the change of $\sigma^{2}$ does not
affect the aforementioned observations of the time-dependent characteristics
of these entropies. The major difference between different values
of $\sigma^{2}$ is that it results in a shift of the values $\overline{S}(z_{c}:z_{e},t)$
and it affects the already slow dispersion rate of $\Psi_{0}$.

\begin{figure}[H]
\includegraphics[width=1\columnwidth]{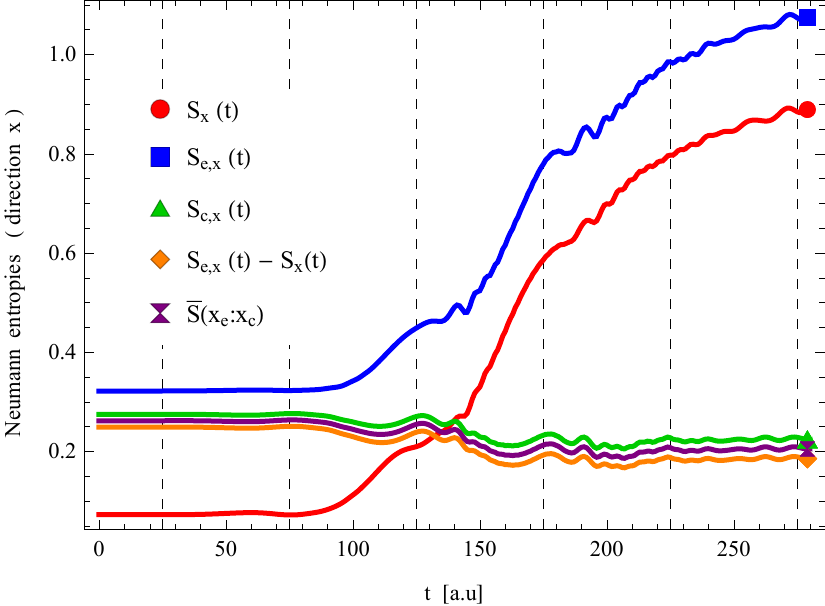}\protect\protect\caption{The time evolution of the various Neumann entropies based on the reduced
density matrices in direction $x$, using the parameters $F=0.1$,
$\mathrm{CEP}=0$. We can see that the $\overline{S}(x_{e}:x_{c},t)$
shares time-dependent features with $\overline{S}(z_{e}:z_{c},t)$,
for example, its maxima are near the zero crossings of the laser pulse.
The major correlation between $x_{e}$ and $x_{c}$ is quantum entanglement. }

\label{fig:sinpulse3_F0.1_NeumEnt_X} 
\end{figure}

The time-dependence of the same of quantum entropies which characterize
the reduced dynamics along the $x$ axis (same along $y$) can be
seen in Fig. \ref{fig:sinpulse3_F0.1_NeumEnt_X}. However, we limited
the range of the time axis (to $280$ atomic time units) in this case,
since one of this entropy calculations is done about $O(N^{4})$ steps
instead of $O(N^{3})$, and it also involves that much interpolation
in order to do integration in Cartesian coordinates.

From Fig. \ref{fig:sinpulse3_F0.1_NeumEnt_X}, we can see a familiar
shape related to the spatial entropy in the form of $S_{x}(t)$, because
the values of the Neumann entropy $S_{x}(t)$ mirrors that of $S_{z}(t)$,
but they are not the same. However, they are actually identical at
$t=0$ due to the spherical symmetry of $1s$ Coulomb state, i.e.
(a single index) tripartite Schmidt decomposition \cite{pati2000tripartiteSchmidt}
of the initial relative wave function exists. Then the laser pulse
causes this wave function to slowly depart from this tripartite Schmidt
state as $S_{x}(t)$ and $S_{z}(t)$ differ more. However, both $S_{x}(t)$
and $S_{z}(t)$ depict the time dependence of spatial entropy adequately.

Now we turn our attention to the particle-particle correlation of
the $x_{e}$ and $x_{c}$ coordinates. First, this correlation is
quantum entanglement because $S_{c,x}(t)$ and its negative conditional
entropy i.e. $S_{e,x}(t)-S_{x}(t)$ stay really close to each other
which is only possible if $x_{e}$ and $x_{c}$ are entangled, therefore
$\overline{S}(x_{e}:x_{c},t)$ is a good entanglement measure. Now,
we can also see that the $\overline{S}(x_{e}:x_{c},t)$ shares some
time-dependent features with $\overline{S}(z_{e}:z_{c},t)$, for example,
its maxima are near the zero crossings of the laser pulse. Note, however,
that the changes in $\overline{S}(x_{e}:x_{c},t)$ are considerably
smaller than those in $\overline{S}(z_{e}:z_{c},t)$. It is somewhat
surprising that there is an overall entanglement decrease in direction
$x$, which we discuss in the next subsection in more detail. This
decrease could be an evidence of the purification between the two
subsystems $x_{e}$ and $x_{c}$, as these coordinates become more
uncorrelated during the physical process.

\begin{figure}[H]
\includegraphics[width=1\columnwidth]{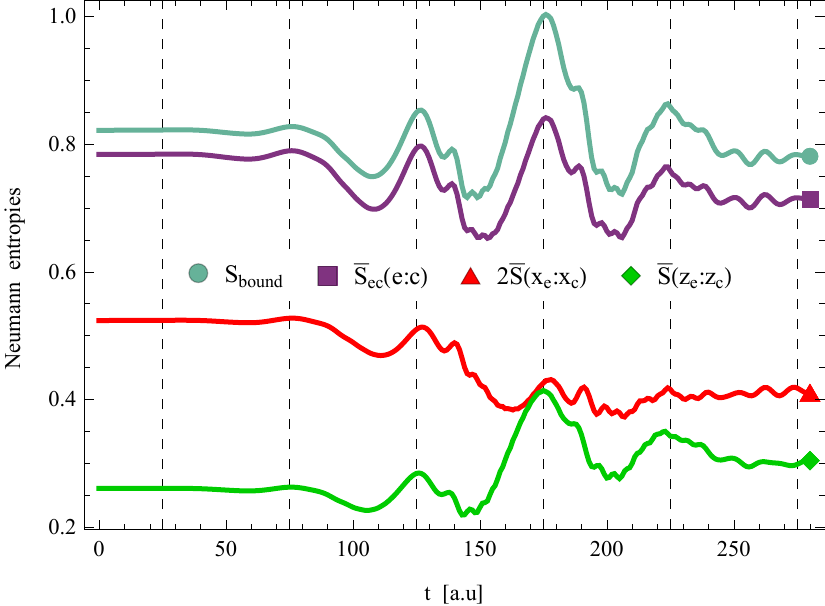}\protect\protect\caption{The time evolution of our electron-core entanglement entropy $\overline{S}_{ec}(e:c,t)$
and the upper bound of the analytic entanglement entropy $S_{\mathrm{bound}}(t)$,
along with time evolution of the directional entropies $2\overline{S}(x_{e}:x_{c},t)$
and $\overline{S}(z_{e}:z_{c},t)$ with parameters $F=0.1$, $\mathrm{CEP}=0$.
The time dependence of $S_{\mathrm{bound}}(t)$ and $\overline{S}_{ec}(e:c,t)$
follow each other with substantial, and slightly increasing gap which
indicates the actual importance of these curves. The total entanglement
entropy reaches a net decrease by the end of the laser pulse. Important
features of $\overline{S}_{ec}(e:c,t)$ are shared with $\overline{S}(z_{e}:z_{c},t)$
(like the correlation with the external electric field, and the definite
positions of the maxima near the zero crossings of the laser field)
which suggests that the relevant physics happens along the polarization
axis. }

\label{fig:sinpulse3_F0.1_MutEnt_All} 
\end{figure}

Finally, in Fig. \ref{fig:sinpulse3_F0.1_MutEnt_All}, we plot the
result of our approximate formula $\overline{S}_{ec}(e:c,t)$ of the
physical core-electron entanglement using (\ref{eq:ent_Sec_mut_avg_formula})
with its analytic upper bound $S_{\mathrm{bound}}(t)$ via (\ref{eq:ent_SN_ec_bound}).
There, we also plot the function $\overline{S}(x_{e}:x_{c},t)+\overline{S}(y_{e}:y_{c},t)=2\overline{S}(x_{e}:x_{c},t)$
and $\overline{S}(z_{e}:z_{c},t)$ for the $z$ subsystem. We see
that our approximate quantification formula $\overline{S}_{ec}(e:c,t)$
is clearly below $S_{\mathrm{bound}}(t)$, with substantial, and slightly
increasing gap. Also the time dependence of these follow each other,
which indicates the actual importance of $\overline{S}_{ec}(e:c,t)$.
It seems to be surprising that the total entanglement entropy shows
a net decrease by the end of the laser pulse, which we will revisit
in the next subsection. This is especially interesting when we take
into account that other important features of the entropy $\overline{S}_{ec}(e:c,t)$
mimic those of the mutual entropy $\overline{S}(z_{e}:z_{c},t)$.
In this sense we could say that the part of the relevant physics happens
along the polarization axis (like the correlation with the external
electric field, and the definite positions of the maxima near the
zero crossings of the laser field) but the perpendicular degrees of
freedom change the overall dynamics of the entanglement from increasing
to decreasing.

We will further explore the dynamics of all types of entanglement
presented so far in the following subsections, while also giving more
insight into the physics, by changing the external field that governs
the process.

\subsection{Parameter dependence of the quantum entropies: electric field strength\label{sub:results_F_dependence}}

In this section we discuss the dependence of the important entanglement
entropies of Section \ref{sec:entanglement-calculation} on the parameter
$F$ i.e. on the strength of the external electric field.

\begin{figure}[H]
\includegraphics[width=1\columnwidth]{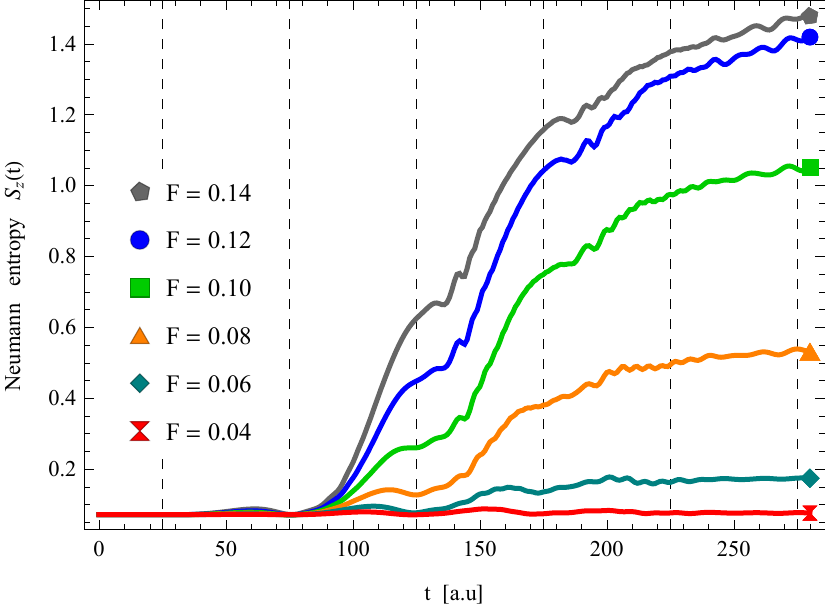}\protect\protect\caption{The time evolution of the Neumann entropy $S_{z}(t)$ for the indicated
values of the parameter $F$, with $\mathrm{CEP}=0$. Below the value
$F=0.04$, we have only a negligible increase in $S_{z}(t)$, i.e.
the relative wave function stays nearly separable in $z$, $\rho$
during the process. This separability quickly breaks down with increasing
$F$. The fast rises of $S_{z}(t)$ are related to the sudden changes
of ionization probability, see Fig. \ref{fig:sinpulse3_F_projection}. }

\label{fig:sinpulse3_F_NeumEnt_Z} 
\end{figure}

In Fig. \ref{fig:sinpulse3_F_NeumEnt_Z} we plot the spatial entropy
$S_{z}(t)$ for the relevant values of $F$ . Comparing these curves
with the ground state population loss of Fig. \ref{fig:sinpulse3_F_projection}
it is easy to correlate the time evolution of $S_{z}(t)$ to the probability
of ionization.

Note that below the value $F=0.04$, we have only a marginal increase
in $S_{z}(t)$, i.e. the relative wave function stays nearly separable
in $z$, $\rho$ during the process. This separability quickly breaks
down with increasing $F$, which is an important information regarding
the applicability of the time-dependent multiconfigurational Hatree
approaches \citep{beck2000mctdhbig} for the simulations of strong
field processes. It is also interesting that we have not found any
specific mark of the tunneling or the over-the-barrier ionization
regimes. Between $F=0.12$ and $F=0.14$, the entropy increase already
slows down as a function of $F$, and one can extrapolate that the
spatial nonseparability has a saturation point near $F=0.14$. We
verified the existence of this maximum value with additional computations.
Therefore, there is a limiting maximal value for $S_{z}(t)$ in the
given time window, which already corresponds also to nearly complete
ionization. The $S_{z}(t)$ is not only the measure of ``spatial
entanglement'', but it is also the total entropy of the $z$ subsystem,
which has consequences regarding the interpretation of the directional
mutual entropies.

\begin{figure}[H]
\includegraphics[width=1\columnwidth]{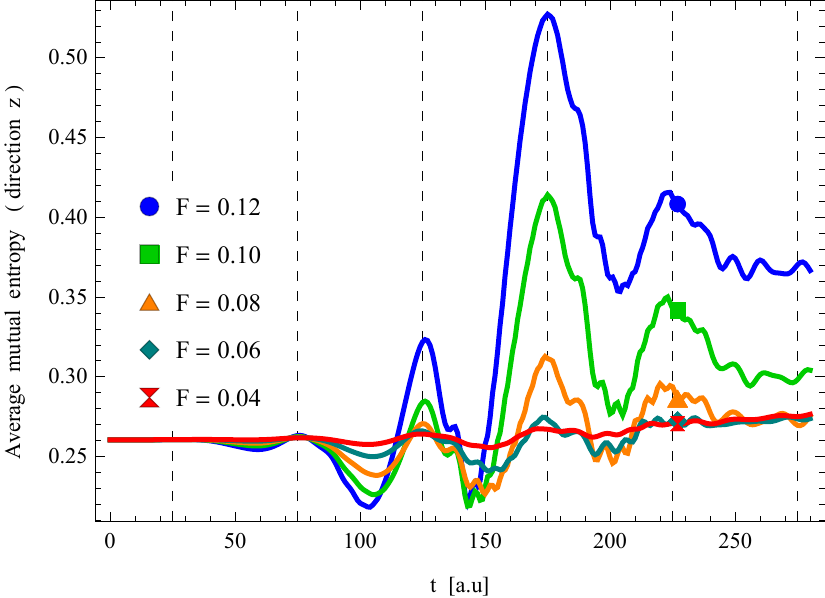}\protect\protect\caption{The time evolution of the mutual entropy $\overline{S}(z_{e}:z_{c},t)$
for the indicated values of the parameter $F$, with $\mathrm{CEP}=0$.
By comparing to Fig. \ref{fig:sinpulse3_F_velocity_z}, we can easily
recognize that the mean relative velocities $\overline{v}_{z}(t)$
are tied to the quantum entanglement in direction $z$. These curves
are very similar to the exact quantum entanglement entropy curves
in Fig. 1. of our former 1D model simulation \cite{czirjak2013rescatterentanglement}.
The local maxima increase with increasing $F$, the largest change
occurring at the main maximum ($t=175$). Positions of the local maxima
almost coincide with the zero crossings of the laser's electric field. }

\label{fig:sinpulse3_F_MutEnt_Z} 
\end{figure}

In Fig. \ref{fig:sinpulse3_F_MutEnt_Z} we plot the average mutual
entropy in the directional $z$ subsystem, $\overline{S}(z_{e}:z_{c},t)$,
for the relevant range of $F$. It is easy to see that the correlation
of this entropy with the shape of the laser pulse becomes more clear
as we increase $F$. The values of the first minima decrease as $F$
increases, but this is reversed for the other local minima. Regarding
the local maxima, they all increase with increasing $F$, the largest
change occurring at the main maximum ($t=175$). Positions of the
local maxima are independent of $F$. We can observe a tunneling regime
feature: the value of this entropy returns to the baseline at the
end of the laser pulse. As the over-the-barrier ionization takes over
( above $F=0.08$) the final value of the entanglement between $z_{e}$
and $z_{c}$ rises with increasing $F$ .

Comparing Fig. \ref{fig:sinpulse3_F_MutEnt_Z} and Fig. \ref{fig:sinpulse3_F_velocity_z},
it is easy to recognize that the mean relative velocities $\overline{v}_{z}(t)$
(or alternatively, momenta) play a particularly important role regarding
quantum entanglement in this direction. During one half cycle of the
laser pulse, as the core and the electron are moving apart, the entanglement
of their respective coordinates $z_{e}$ and $z_{c}$ increases proportionally
to the magnitudes of their relative velocities. The value of their
entanglement decreases when deceleration occurs, and reaches its minimum
value when the particles' relative motion stops. The final value of
entanglement is also related to this velocity.

The results presented in Fig. \ref{fig:sinpulse3_F_MutEnt_Z} are
even more interesting if we compare them to the exact quantum entanglement
entropy curves in Fig. 1. of our former 1D model simulation \citep{czirjak2013rescatterentanglement}.
Despite that the average mutual entropy $\overline{S}(z_{e}:z_{c},t)$
includes an increasing ``background'' (since the composite system
is always in a mixed state in the 3D model), the main features of
the temporal dependence in Fig. \ref{fig:sinpulse3_F_MutEnt_Z} and
in Fig. 1. of \citep{czirjak2013rescatterentanglement} exhibit a
very good qualitative agreement: the position of the local maxima
coincide with the zeros of the laser pulse, the main maximum of the
entropy is roughly the double of its initial value, and the asymptotic
value at the end of the simulation time scales roughly the same way
to the corresponding maximum values. This agreement strongly supports
our opinion that the average mutual entropy $\overline{S}(z_{e}:z_{c},t)$
is a useful measure of quantum entanglement for the degrees of freedom
along the direction of the laser polarization in the 3D case. The
agreement also justifies the use of the delta-potential in the 1D
simulation, because the resulting exact core-electron quantum entanglement
quantitatively correctly describes the corresponding entanglement
dynamics of the 3D case.

\begin{figure}[H]
\includegraphics[width=1\columnwidth]{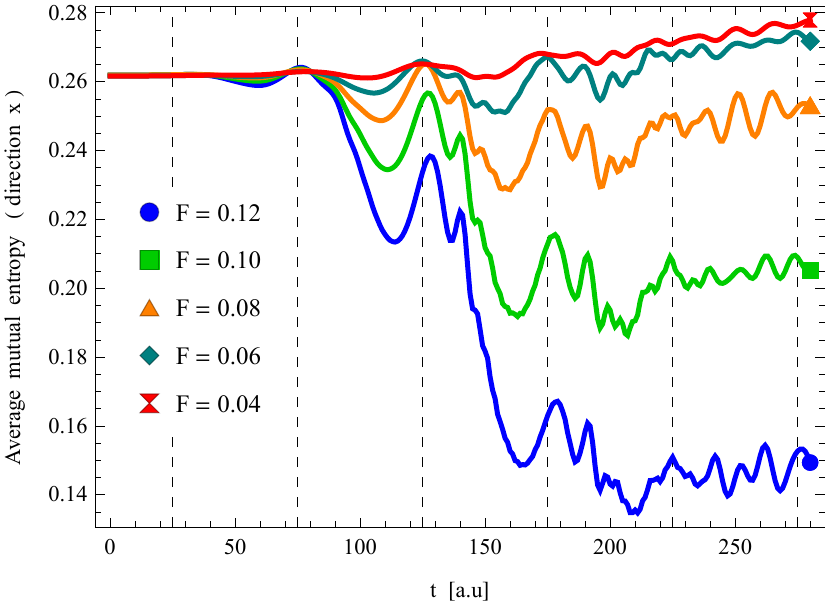}\protect\protect\caption{The time evolution of the mutual entropy $\overline{S}(x_{e}:x_{c},t)$
for the indicated values of the parameter $F$, with $\mathrm{CEP}=0$.
This figure shows more clearly the striking feature that was already
present in Fig. \ref{fig:sinpulse3_F0.1_NeumEnt_X}: the average mutual
entropy in the direction $x$ decreases surprisingly strongly with
increasing $F$ in the over-the-barrier ionization regime. }

\label{fig:sinpulse3_F_MutEnt_X} 
\end{figure}

Regarding the transverse direction $x$, first we note that the time
dependence of $S_{x}(t)$ is very similar to that of $S_{z}(t)$ and
it scales with $F$ also in an analogous way, therefore we do not
plot $S_{x}(t)$. We plot $\overline{S}(x_{e}:x_{c},t)$ in Fig. \ref{fig:sinpulse3_F_MutEnt_X}
in analogy to Fig. \ref{fig:sinpulse3_F_MutEnt_Z}. This figure shows
more clearly the striking feature that was already present in Fig.
\ref{fig:sinpulse3_F0.1_NeumEnt_X}: the average mutual entropy in
the transverse direction decreases surprisingly strongly with increasing
$F$ in the over-the-barrier ionization regime. This unexpected behavior
is of purely quantum mechanical nature, contrary to direction $z$:
since $\overline{v}_{x}(t)=0$, there is no ``classical\textquotedblright \ explanation
based on the Ehrenfest kinematics. However, the positions of the local
maxima $\overline{S}(x_{e}:x_{c},t)$ are still tied to the zero crossings
of the laser field. There is an importance of the tunneling regime
($F=0.0624$ and below), where the average mutual entropies $\overline{S}(x_{e}:x_{c},t)$
and $\overline{S}(z_{e}:z_{c},t)$ have almost the same overall behavior
and show an entropy increase.

\begin{figure}[H]
\includegraphics[width=1\columnwidth]{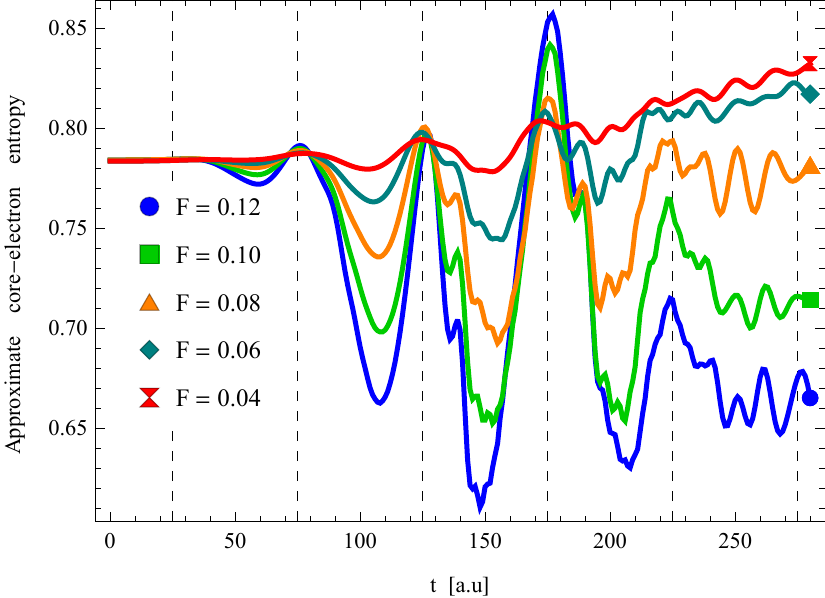}\protect\protect\caption{The time evolution of the approximate core-electron entropy $\overline{S}_{ec}(e:c,t)$
for the indicated values of the parameter $F$, with $\mathrm{CEP}=0$.
Due to its construction, it inherits its features from $\overline{S}(z_{e}:z_{c},t)$
and $\overline{S}(x_{e}:x_{c},t)$. Surprisingly, the entropy decrease
of the transverse directions dominate the entropy increase in direction
$z$, therefore this approximate core-electron entanglement decreases
with increasing $F$ in the over-the-barrier ionization regime.}

\label{fig:sinpulse3_F_MutEnt_All} 
\end{figure}

Finally, in Fig. \ref{fig:sinpulse3_F_MutEnt_All}, we plot the approximate
core-electron entanglement $\overline{S}_{ec}(e:c,t)$, defined in
Eq. (\ref{eq:ent_Sec_mut_avg_formula}). Due to its construction,
it inherits its features from $\overline{S}(z_{e}:z_{c},t)$ and $\overline{S}(x_{e}:x_{c},t)$
in the following way: if the value of $F$ ensures pure tunnel ionization,
then $\overline{S}_{ec}(e:c,t)$ gains a net increase by the end of
the laser pulse, otherwise the core-electron entanglement decreases
with increasing $F$, which is a rather surprising result. Other important
features of $\overline{S}(z_{e}:z_{c},t)$ are preserved also for
$\overline{S}_{ec}(e:c,t)$: the presence of the local maxima at the
zero crossings of the laser field, the general nature of the correlations,
and its link to the mean velocity.

\subsection{Parameter dependence of the quantum entropies: carrier envelope phase\label{sub:results_CEP_dependence}}

In this section we investigate the effects of the carrier envelope
phase ($\mathrm{CEP)}$ on the process.

\begin{figure}[H]
\includegraphics[width=1\columnwidth]{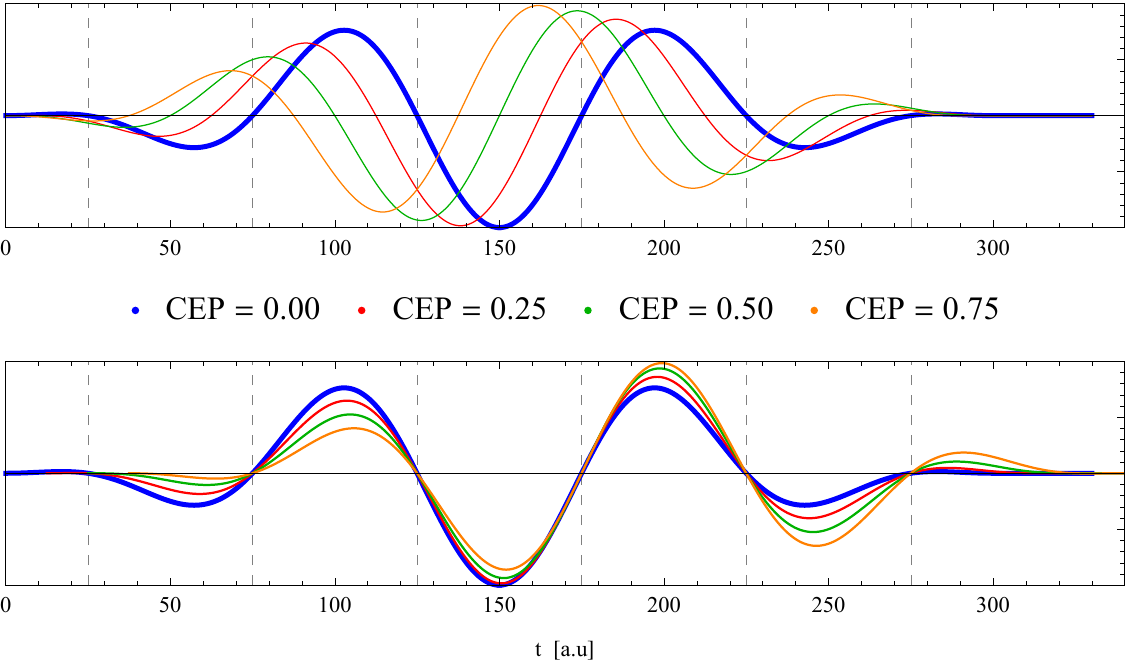}\protect\protect\caption{Plots of the laser pulses' electric fields $E_{z}(t)$ versus time
with four selected values of the parameter $\mathrm{CEP}$, where
the thick blue curves indicate the case of $\mathrm{CEP=}0$. The
vertical axes range from $-F$ to $F$ and represent the strength
of the electric field. Plots in the upper panel are according to the
formula \eqref{eq:sim_sinpulse3_E} then we applied a $\mathrm{CEP}$
dependent shift in time to make the zero crossings coincide (lower
panel). We plot the time-dependence of some selected quantities in
Figures \ref{fig:sinpulse3_CEP_velocity_z} to \ref{fig:sinpulse3_CEP_MutEnt_All}
with this shift applied. }

\label{fig:sinpulse3_CEP_Pulse} 
\end{figure}

In the upper panel of Fig. \ref{fig:sinpulse3_CEP_Pulse} we plot
the electric field of the laser pulse for our selected $\mathrm{CEP}$
values, with the strength of the electric field parameter set to $F=0.1$.
For the sake of better comparability, we apply the following $\mathrm{CEP}$
dependent transformation in time: we shift backwards the time domains
in the case of nonzero $\mathrm{CEP}$ values such that the zero crossings
of the various laser pulses coincide, as shown in the lower panel
of Fig. \ref{fig:sinpulse3_CEP_Pulse}. We plot the time dependence
of some selected quantities in the following figures with this shift
applied.

\begin{figure}[H]
\includegraphics[width=1\columnwidth]{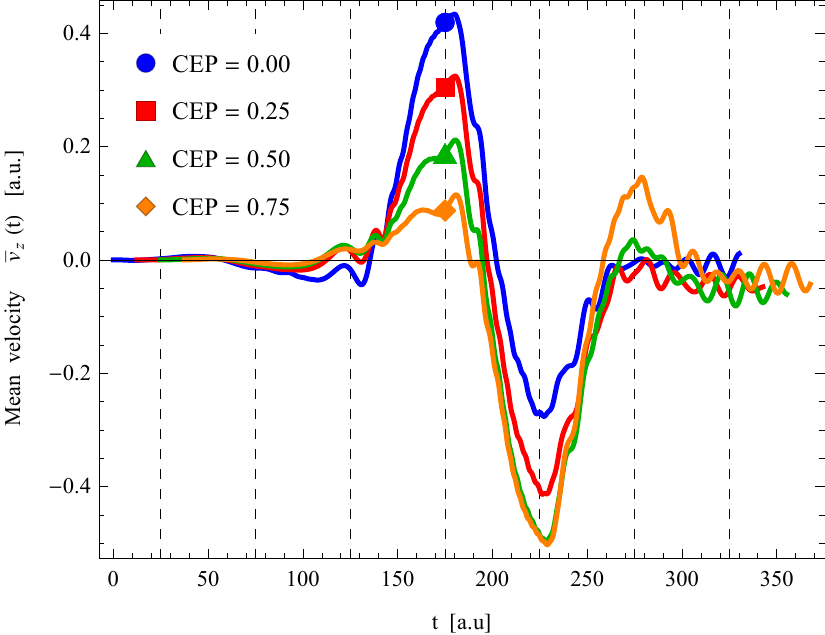}\protect\protect\caption{Plots of the mean velocity $\overline{v}_{z}(t)$ of the relative
wave function versus time for the indicated $\mathrm{CEP}$ parameters,
with $F=0.1$. }

\label{fig:sinpulse3_CEP_velocity_z} 
\end{figure}

\begin{figure}[H]
\includegraphics[width=1\columnwidth]{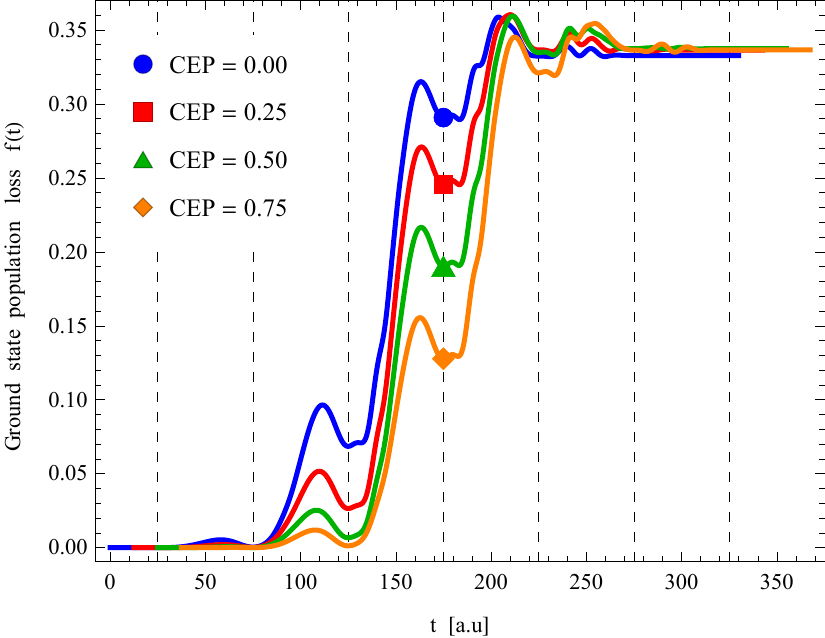}\protect\protect\caption{Plots of the ground state population loss of the relative wave function
versus time for the indicated $\mathrm{CEP}$ parameters, with $F=0.1$. }

\label{fig:sinpulse3_CEP_projection-1} 
\end{figure}

We plot the $\mathrm{CEP}$ dependence of the ground state population
loss in Fig. \ref{fig:sinpulse3_CEP_projection-1} and the mean velocity
$\overline{v}_{z}(t)$ in Fig. \ref{fig:sinpulse3_CEP_velocity_z}
using the above mentioned transformation. For each $\mathrm{CEP}$
value, the dynamical properties of the system stay synchronized to
the local minima, maxima and zero crossings of the laser pulses. The
values of the ground state population loss at the end of the laser
pulse are nearly unaffected by the parameter $\mathrm{CEP}$. The
corresponding values of $\overline{v}_{z}(t)$ are only slightly affected
by the $\mathrm{CEP}$ change.

\begin{figure}[H]
\includegraphics[width=1\columnwidth]{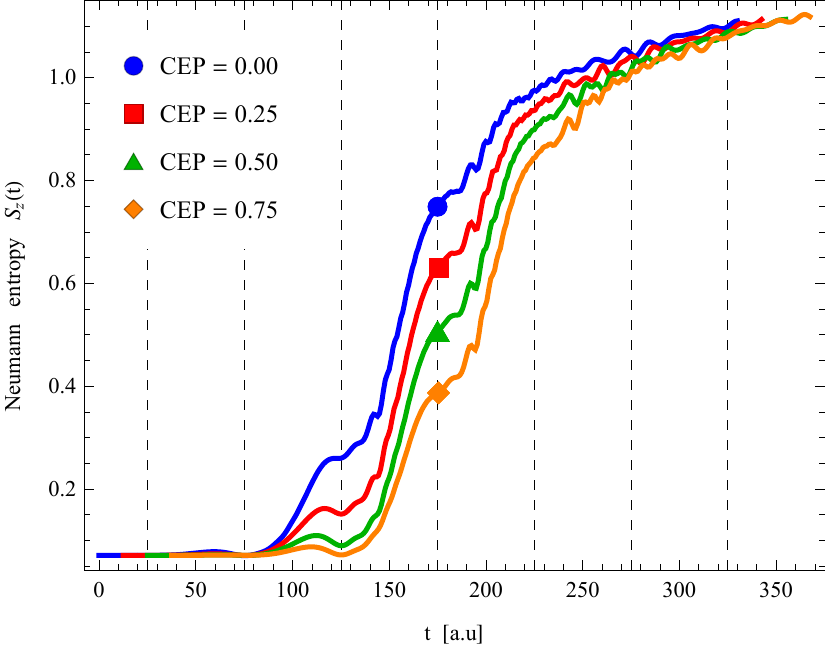}\protect\protect\caption{Plots of the Neumann entropy $S_{z}(t)$ versus time for the indicated
$\mathrm{CEP}$ parameters, with $F=0.1$. }

\label{fig:sinpulse3_CEP_NeumEnt_Z} 
\end{figure}

\begin{figure}[H]
\includegraphics[width=1\columnwidth]{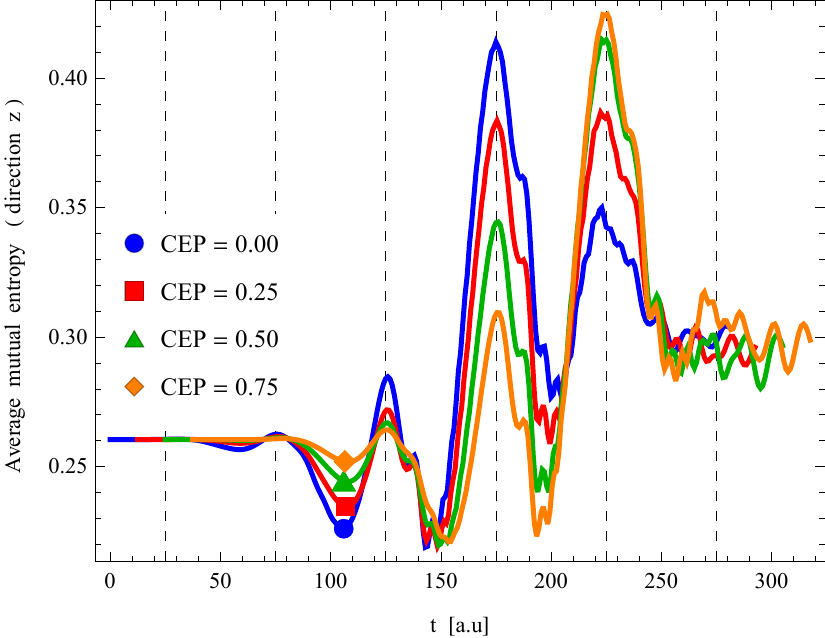}\protect\protect\caption{The plot of the mutual entropy $\overline{S}(z_{e}:z_{c},t)$ for
the indicated $\mathrm{CEP}$ parameters, with $F=0.1$. Note that
the peak at $t=175$ shrinks as the $\mathrm{CEP}$ increases while
the peak at $t=225$ increases. }

\label{fig:sinpulse3_CEP_MutEnt_Z} 
\end{figure}

\begin{figure}[H]
\includegraphics[width=1\columnwidth]{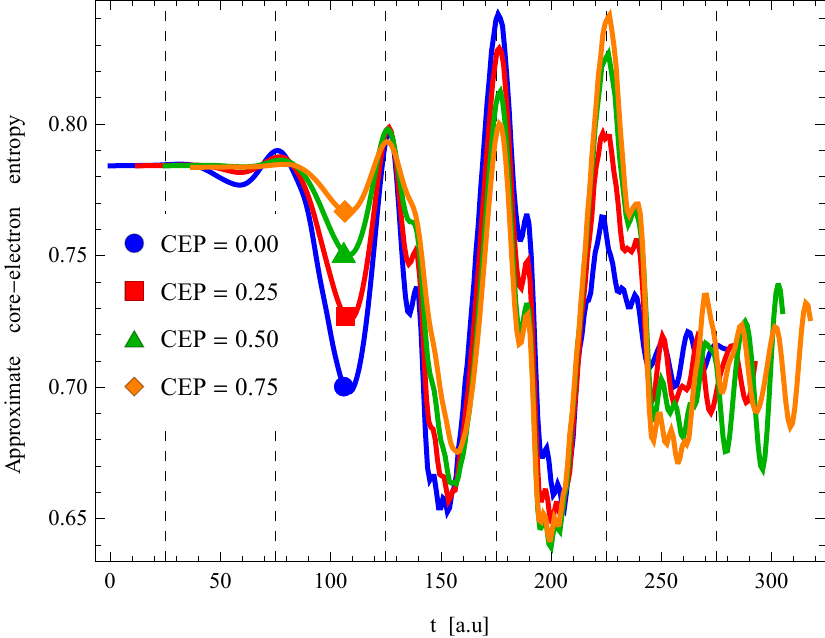}\protect\protect\caption{The plot of the electron-core entropy $\overline{S}_{ec}(e:c,t)$
for the indicated $\mathrm{CEP}$ parameters, with $F=0.1$. }

\label{fig:sinpulse3_CEP_MutEnt_All} 
\end{figure}

The entanglement properties of the system inherit the above $\mathrm{CEP}$
related features. To show this, we plot the $\mathrm{CEP}$ dependence
of the entropy of the ``spatial entanglement'' in Fig. \ref{fig:sinpulse3_CEP_NeumEnt_Z},
the entropy of nonseparability in direction $z$ in Fig. \ref{fig:sinpulse3_CEP_MutEnt_Z},
and our approximated core-electron entanglement entropy in Fig. \ref{fig:sinpulse3_CEP_MutEnt_All}
including already the $\mathrm{CEP}$ dependence in direction $x$.
In the latter two Figures, we can see that the local maxima still
coincide with the zeros of the electric fields, independently of the
$\mathrm{CEP}$ values, and the $\mathrm{CEP}$ has barely any effect
on the final values. However, the actual values of the ionization,
the velocities and all the entropies change considerably with respect
to each other between subsequent half cycles, depending on the value
of the $\mathrm{CEP}$ parameter. For example, in Fig. \ref{fig:sinpulse3_CEP_NeumEnt_Z}
the peak at $t=175$ shrinks as $\mathrm{CEP}$ increases and the
peak value at $t=225$ grows synchronously. We have found it interesting
that the latter entropy acquires its largest value near $\mathrm{CEP}=0.75$
and not $\mathrm{CEP}=0.0$, where we have the largest value of $E_{z}(t)$.
Thus, although the parameter ${\rm CEP}$ changes the sub-cycle dynamics
of both these entropies considerably, its value does not affect our
main observations about the overall time-dependent entropy dynamics.

\section{Summary\label{sec:summary}}

In this paper, we applied the theory of quantum entanglement and the
concepts of quantum information theory to describe the time-dependent
correlation properties of an electron and its parent ion-core under
the influence of an external laser pulse which is strong enough to
liberate the electron by tunnel or by over-the-barrier ionization.
The computation of the standard entanglement measure i.e. the Neumann
entropy of either the electron or the core density matrix for this
problem is numerically prohibitive in its full dimensionality, therefore
we choose to partition the interacting system along the spatial directions
parallel and perpendicular to the laser polarization axis, denoted
by $z$ and $x$, respectively. These direction-wise reduced dynamics
still retain all pair correlations in $x$ and $z$. To analyze the
corresponding pair correlations between the electron and the ion-core
coordinates, we used several kinds of Neumann entropies that can be
calculated from the one-dimensional density matrices of the system.
Based on the concepts of quantum conditional entropy and quantum mutual
entropy, we introduced average mutual entropies between the electron's
and the ion-core's spatial position along the $x$ and $z$ directions
as suitable and useful correlation measures. We constructed an approximate
formula, Eq. \eqref{eq:ent_Sec_mut_avg_formula}, to quantify the
total particle-particle entanglement between the electron and the
ion-core, based on the direction-wise mutual entropies.

We analyzed the nature of the correlations in each direction and we
found that they are based on the same fundamental features of this
system. For example in direction $z$, the ion-core entropy $S_{c,z}(t)$
behaves like a correlation entropy, because the ion-core density matrix
is close to that of the center of mass which has zero entropy. The
spatial entropy $S_{z}(t)$ is concentrated in the direction-wise
electron entropy $S_{e,z}(t)$, which also incorporates a correlation
part. The resulting $S_{e,z}(t)-S_{z}(t)$, which is the negative
conditional entropy of the ion-core, becomes positive and has many
features in common with $S_{c,z}(t)$. In most of the simulations,
these two stay really close to each other, which means that the state
as the function of the $z_{e}$ and $z_{c}$ coordinates shows dominantly
quantum entanglement. The same is true with respect to the $x_{e}$
and $x_{c}$ coordinates. This behavior is very different from pure
state entanglement, because these directional subsystems are in mixed
states.

We analyzed the correlation entropy relations in each direction and
we found that the zero crossings of the electric field almost coincide
with their local maxima. These results in direction $z$ are also
in a good agreement with our earlier one dimensional simulations.
The correlations along the $x$ and $z$ directions are very similar
to each other if the process stays in the tunnel ionization regime.
In the over-the-barrier ionization regime, we found entropy increase
along $z$ but a surprising entropy decrease in the transverse directions
which makes also the total core-electron entanglement entropy to decrease,
contrary to what we expected.

We investigated the dependence of these proposed measures of entanglement
dynamics on the strength and the carrier-envelope phase of the driving
laser pulse. We found many features of quantum entropies that do not
depend on these parameters, like the electron-core entanglement has
local maxima always near the zero crossings of the laser pulse. We
found that while the intensity of the field governs the dynamics as
a whole, the carrier envelope phase changes the sub-cycle dynamics
of the strong field ionization.

Based on our simulations, we also calculated some relevant quantities
that contribute to the physical picture of strong field ionization.
We found that the ground state of the simulated relative wave function
is almost separable, and it remains so if the field is weak. The loss
of the ground state population is a good measure of ionization, and
that the net effect of the ionized waves results in a mean velocity
$v_{z}(t)$ which is more and more similar to the corresponding motion
of a classical electron as the laser intensity increases, apart from
the nonzero final velocity.

We think that our results will be useful regarding the interpretation
of quantum measurements, especially in connection with strong-field
processes, using e.g. COLTRIMS or other reaction microscopes \cite{dorner2000COLTRIMS,ullrich2003COLTRIMS}.
An obvious but not trivial extension of our present work could be
the calculation of electron entanglement in double ionization \cite{becker2012correlation}.
We also hope to inspire further developments in quantum information
theory.
\begin{acknowledgments}
The authors thank W.\ Becker, P.\ Földi, K.\ Varjú and S.\ Varró
for stimulating discussions. Szilárd Majorosi was supported by the
project GINOP-2.3.2-15-2016-00036 of the Ministry of National Economy
of Hungary. Partial support by the ELI-ALPS project is also acknowledged.
The ELI-ALPS project (GOP-1.1.1-12/B-2012-000, GINOP-2.3.6-15-2015-00001)
is supported by the European Union and co-financed by the European
Regional Development Fund. 
\end{acknowledgments}

\appendix
%dummy comment inserted by tex2lyx to ensure that this paragraph is not empty

\section{Quantification of bipartite quantum entanglement\label{sec:quantum-entanglement}}

\subsection{Schmidt decomposition and entanglement}

In this appendix we recall the standard theory of quantum entanglement
for bipartite systems emphasizing the features specific to states
described by square integrable coordinate wave-functions of infinite
dimensional Hilbert spaces. In our problem the two parts $e$ and
$c$, are two distinguishable particles, the electron and its parent
ion-core. The composite system $ec$ is assumed to be a closed quantum
system in a pure state represented by the wave function $\Psi_{ec}(\mathbf{r}_{e},\mathbf{r}_{c},t)$.
The two subsystems are entangled if $\Psi_{ec}$ is not separable
with respect to the coordinates of these subsystems: 
\begin{equation}
\Psi_{ec}(\mathbf{r}_{e},\mathbf{r}_{c},t)\neq\Psi_{e}(\mathbf{r}_{e},t)\Psi_{c}(\mathbf{r}_{c},t).
\end{equation}
It is well-known that then the result of the measurement of subsystem
$e$ affects the outcome of measurements on subsystem $c$ and vice-versa.
That is, performing measurement on either particle changes the other
particle's quantum state in a nonlocal manner.

To quantify the entanglement, we need the relevant concept of density
matrices. The composite system is described by the two-particle pure
state density matrix: 
\begin{equation}
\varrho_{ec}(\mathbf{r}_{e}^{\prime},\mathbf{r}_{e},\mathbf{r}_{c}^{\prime},\mathbf{r}_{c},t)=\Psi_{ec}^{\ast}(\mathbf{r}_{e}^{\prime},\mathbf{r}_{c}^{\prime},t)\Psi_{ec}(\mathbf{r}_{e},\mathbf{r}_{c},t).\label{eq:ent_dm_two_particle}
\end{equation}
and the single particle density matrices\ are obtained by tracing
over the other particle's degrees of freedom. The reduced single particle
core density matrix is 
\begin{equation}
\varrho_{c}(\mathbf{r}_{c}^{\prime},\mathbf{r}_{c},t)=\mathrm{Tr}_{e}\left[\hat{\varrho}_{ec}\right]=\int\varrho_{ec}(\mathbf{r}_{e},\mathbf{r}_{e},\mathbf{r}_{c}^{\prime},\mathbf{r}_{c},t){\rm d}\mathbf{r}_{e}^{3}\label{eq:ent_dm_core}
\end{equation}
and the reduced single particle electron density matrix is 
\begin{equation}
\varrho_{e}(\mathbf{r}_{e}^{\prime},\mathbf{r}_{e},t)=\mathrm{Tr}_{c}\left[\hat{\varrho}_{ec}\right]=\int\varrho_{ec}(\mathbf{r}_{e}^{\prime},\mathbf{r}_{e},\mathbf{r}_{c},\mathbf{r}_{c},t){\rm d}\mathbf{r}_{c}^{3}.\label{eq:ent_dm_electron}
\end{equation}
These quantities contain every quantum information about the respective
single particle properties, and they are directly related to the entanglement
information we need. To show this, we refer to the Schmidt theorem
\citep{eberly2006entschmidt,luo2003entanglement}, which states that
there exists a unique decomposition of the entangled wavefunction
$\Psi_{ec}$ of the bipartite system $ec$ into a sum of the following
form: 
\begin{equation}
\Psi_{ec}(\mathbf{r}_{e},\mathbf{r}_{c},t)=\sum\limits _{k}\sqrt{\lambda_{k}(t)}\phi_{k}(\mathbf{r}_{c},t)\psi_{k}(\mathbf{r}_{e},t).\label{eq:ent_schmidt_decomp}
\end{equation}
where $\phi_{k}(\mathbf{r}_{c},t)$ and $\psi_{k}(\mathbf{r}_{e},t)$
are orthonormal basis functions in the respective spaces. They are
acquired after the diagonalization of the single particle reduced
density matrices (\ref{eq:ent_dm_core}) and (\ref{eq:ent_dm_electron})
as 
\begin{equation}
\varrho_{e}(\mathbf{r}_{e}^{\prime},\mathbf{r}_{e},t)=\sum\limits _{k}\lambda_{k}(t)\psi_{k}^{\ast}(\mathbf{r}_{e}^{\prime},t)\psi_{k}(\mathbf{r}_{e},t),\label{eq:ent_dm_diag_electron}
\end{equation}
\begin{equation}
\varrho_{c}(\mathbf{r}_{c}^{\prime},\mathbf{r}_{c},t)=\sum\limits _{k}\lambda_{k}(t)\phi_{k}^{\ast}(\mathbf{r}_{c}^{\prime},t)\phi_{k}(\mathbf{r}_{c},t),\label{eq:ent_dm_diag_core}
\end{equation}
i.e the formula (\ref{eq:ent_schmidt_decomp}) contains the eigenvectors
$\phi_{k}$, $\psi_{k}$ as the Schmidt basis functions, and the countably
many common eigenvalues $\lambda_{k}(t)$ of $\varrho_{e}$ and $\varrho_{c}$
density matrices respectively. We note that in this continuous variable
case the diagonalization of (\ref{eq:ent_dm_core}) or (\ref{eq:ent_dm_electron})
actually involves the solution of a homogenous Fredholm integral equation
of the second kind. In addition -- contrary to discrete variable systems
-- these density matrices are usually highly singular, due to the
trace condition Tr$\varrho_{e}$=Tr$\varrho_{c}=1$ they contain infinitely
many zero or close to zero eigenvalues. Therefore, it is necessary
to introduce an ordering of the eigenvalues $\lambda_{1}\geq\lambda_{2}\geq\lambda_{3}\geq\ldots$
and then to use only a finite number of them which are greater than
an adequately small threshold number $\epsilon$.

The eigenvalues $\lambda_{k}(t)$ \ allow one to quantify the entanglement
of the particles (subsystems) $e$ and $c$ by introducing quantum
entropies \citep{eltschka2014entquantifying,lin2014enthydrogen}.
Most frequently we use here the von Neumann entropy 
\begin{equation}
S_{N}(t)=-\mbox{Tr}\left[\hat{\varrho}_{c}(t)\ln\hat{\varrho}_{c}(t)\right]=-\sum\limits _{k}\lambda_{k}(t)\ln\lambda_{k}(t),\label{eq:ent_ent_neumann}
\end{equation}
and in certain cases the linear entropy 
\begin{equation}
S_{L}(t)=1-\mbox{Tr}\left[\hat{\varrho}_{c}^{2}(t)\right]=1-\sum\limits _{k}\lambda_{k}^{2}(t).\label{eq:ent_ent_linear}
\end{equation}
The von Neumann entropy obeys some natural requirements, and it also
has a quantum information theoretic appeal \citep{schumacher1995quantumcoding}
while the linear entropy (\ref{eq:ent_ent_linear}) is easier to calculate,
since diagonalization is not necessary. However, both of these entropies
generally tend to behave the same way in this simple bipartite configuration:
if a subsystem is in a pure state they assume the value $0$, and
they increase as the ``mixedness\textquotedblright{} of the subsystem's
state increases . It is important that this quantification does not
straightforwardly generalize to the case where the composite system
is divided into more than two subsystems \cite{carteret2000schmidtgeneral}.

For independent systems the total density operator is the tensorial
product of those of the subsystems and then the Neumann entropy of
the composite system is exactly the sum of the Neumann entropies of
the subsystems. In our case, however, when by the very nature of the
problem $e$ and $c$ are not independent, only strong subadditivity
holds \citep{wehrl1978entropyproperties}, which gives an upper bound
of the composite system's entropy as 
\begin{equation}
S_{N}(\varrho_{ec})\leq S_{N}(\varrho_{e})+S_{N}(\varrho_{c}),\label{eq:qinf_SN_subadd}
\end{equation}
A useful lower bound is given by Araki-Lieb inequality as 
\begin{equation}
\left\vert S_{N}(\varrho_{e})-S_{N}(\varrho_{c})\right\vert \leq S_{N}(\varrho_{ec}).\label{eq:qinf_SN_Araki_Lieb}
\end{equation}

\subsection{Correlation types and quantum information}

In general, $\varrho_{ec}$ involves both classical and quantum correlations.
Then it is crucial to recognize the features of these, and to do that,
we recall their meaning first. If a bipartite system contains only
classical correlations between the two subsystems, then it has a density
matrix of the following form: 
\begin{equation}
\varrho_{ec}^{\mathrm{(cl)}}=\sum_{k}w_{k}\cdot\varrho_{e}^{(k)}\otimes\varrho_{c}^{(k)},\label{eq:qinf_dmz_cl}
\end{equation}
where $w_{k}$ satisfy $\sum_{k}w_{k}=1$ and $w_{k}\geq0$. We are
dealing with some form of quantum entanglement only if the density
matrix of the system does not satisfy (\ref{eq:qinf_dmz_cl}). We
denote the corresponding class of nonclassical density matrices generally
as $\varrho_{ec}^{({\rm quant})}$. A special case of this is the
entangled pure state density matrix $\varrho_{ec}^{{\rm (pure})}$
defined in \eqref{eq:ent_dm_two_particle} which will serve as an
important analytic example for quantum entanglement.

In the following, we recall relevant entropic quantities of quantum
information theory that suit the task of determination and quantification
of entanglement. We will denote the composite system by $EC$, and
its subsystems by $E$ and $C$. We also simplify the notation of
the entropies as $S(EC)=S_{N}(\varrho_{ec})$, $S(E)=S_{N}(\varrho_{e})$,
$S(C)=S_{N}(\varrho_{c})$.

\subsection{Quantum conditional entropy}

The quantum conditional entropy corresponding to a subsystem can be
introduced based on the conditional density or amplitude operator
\citep{cerf1997quantuminf,cerf1999quantcondprob}, but we consider
the following formula for the definition 
\begin{equation}
S(E|C)=S(EC)-S(C)\label{eq:qinf_ent_conditional}
\end{equation}
for the quantum conditional entropy of subsystem $E$, and $S(C|E)$
is the quantum conditional entropy of subsystem $C$. This characterizes
the remaining entropy or information of $E$ after $C$ has been measured
completely. Both quantum conditional entropies can generally be interpreted
the same way as the classical ones, but they can have negative values.
They behave exactly the same way for classical correlations as their
classical counterparts: they are nonnegative 
\begin{equation}
\varrho_{ec}^{\mathrm{(cl)}}\Longrightarrow S(E|C)\geq0\text{ and }S(C|E)\geq0.\label{eq:qinf_ent_cond_cl}
\end{equation}
However, when either of them is negative, 
\begin{equation}
S(E|C)\leq0\text{ or }S(C|E)\leq0\Longrightarrow\varrho_{ec}^{\mathrm{(quant)}}\label{eq:qinf_ent_cond_quant}
\end{equation}
then the composite system is entangled, which leads e.g. to a violation
of the Bell inequalities. Note that the converses of (\ref{eq:qinf_ent_cond_cl})
and (\ref{eq:qinf_ent_cond_quant}) are not true and also $S(E)-S(EC)$
is positive in case of quantum entanglement. For example, in case
of pure composite systems we have 
\begin{equation}
\varrho_{ec}^{\mathrm{(pure)}}\Longrightarrow S(E|C)=-S(C)=-S(E).\label{eq:qinf_ent_cond_pure}
\end{equation}
and $S(C)=S(E)$ is positive. Because of this, quantum entanglement
is sometimes called ``supercorrelation\textquotedblright \ and introduces
virtual information which describes that the measurement changes the
quantum state of the other subsystem.

\subsection{Quantum mutual entropy}

Quantum mutual entropy is the shared entropy or shared information
between subsystems $E$ and $C$. It can be defined using a mutual
density or amplitude operator \citep{cerf1997quantuminf}, but we
use the definition 
\begin{equation}
S(E:C)=S(E)+S(C)-S(EC).\label{eq:qinf_ent_mutual}
\end{equation}
It can be also interpreted as the decrease of entropy of subsystem
$E$ due to the knowledge of $C$ (and vice-versa). Because of this,
we note that the conditional entropy and mutual entropy are related
in the respective subsystems as 
\begin{equation}
S(E:C)=S(E)-S(E|C).\label{eq:qinf_ent_complement}
\end{equation}
The quantum mutual entropy is by construction symmetric and its values
are always nonnegative. For classical correlations: 
\begin{equation}
\varrho_{ec}^{\mathrm{(cl)}}\Longrightarrow S(E:C)\leq\min\left[S(E),S(C)\right].\label{eq:qinf_ent_mut_cl}
\end{equation}
If the values of $S(E:C)$ extend above this classical limit then
there is quantum entanglement between $E$ and $C$: 
\begin{equation}
\min\left[S(E),S(C)\right]\leq S(E:C)\Longrightarrow\varrho_{ec}^{\mathrm{(quant)}}.\label{eq:qinf_ent_mut_quant}
\end{equation}
Unfortunately again, it is not true that below the classical limit
(\ref{eq:qinf_ent_mut_cl}) there could not be quantum effects between
the two subsystems. The upper limit of the quantum mutual entropy
is 
\begin{equation}
S(E:C)\leq2\min\left[S(E),S(C)\right],\label{eq:qinf_ent_mut_limit}
\end{equation}
which can be derived from the Araki-Lieb inequality (\ref{eq:qinf_SN_Araki_Lieb}).

It is instructive to observe that for pure state composite systems,
like EPR pairs, $S(E:C)$ is at the upper limit: 
\begin{equation}
\varrho_{ec}^{\mathrm{(pure)}}\Longrightarrow S(E:C)=S(E)+S(C)=2S(E).\label{eq:qinf_ent_pure}
\end{equation}
Based on this and using the exactness of (\ref{eq:qinf_ent_pure}),
a unified entanglement or quantum nonseparability measure can be defined
which we denote as the average mutual entropy: 
\begin{equation}
\overline{S}(E:C)=\frac{1}{2}S(E:C)\label{eq:qinf_ent_mut_entang}
\end{equation}
which is the same as (\ref{eq:ent_ent_neumann}) in pure bipartite
quantum systems. We can also use this to deduce whether we are dealing
with entanglement: if we are near the limit (\ref{eq:qinf_ent_mut_limit}),
i.e. $\overline{S}$ is close to $\min\left[S(E),S(C)\right]$, then
entanglement is the major correlation. The formulae (\ref{eq:qinf_ent_mutual}),
(\ref{eq:qinf_ent_conditional}), (\ref{eq:qinf_ent_mut_entang})
can be used for the analysis of the entanglement dynamics of the directional
bipartite subsystems of (\ref{eq:ent_dm_two_particle}). But we have
to be careful because (\ref{eq:qinf_ent_mut_entang}) is a general
measure of correlations and entanglement e.g. nonseparability, and
does not imply entanglement under general conditions.

\bibliographystyle{unsrtnat}
\bibliography{0Bibliography,2Bibliography}

\clearpage{}

\begin{comment}
The appendix ends here, these figures belong to Section \ref{sec:results}
of the main text. 
\end{comment}

\pagebreak{}

\pagebreak{}

\pagebreak{}

\pagebreak{}

\end{document}